\documentstyle[12pt,aaspp4,amssym]{article} 

\begin{document}

%
%
%
\newcommand{\app}{$\sim$}
\newcommand{\prop}{$\propto$}
\newcommand{\tento}[1]{$10^{#1}$}
\newcommand{\vzs}{$\times$}
\newcommand{\lb}{$\lambda$}
\newcommand{\mm}{$\pm$}
\newcommand{\smle}{$\ddot{\smile}$}
%
%
\newcommand{\lsol}{$L_\odot$}
\newcommand{\msol}{M$_\odot$}
\newcommand{\ebv}{\hbox{$E(B\!-\!V)$}}
\newcommand{\hbc}{H$_\circ$}
\newcommand{\lbol}{$L_{bol}$}
\newcommand{\mdot}{$\dot{\rm M}$}
\newcommand{\acr}{\msol\,yr$^{-1}$}
%
%
\newcommand{\kms}{km\,s$^{-1}$}
\newcommand{\ergs}{ergs\,s$^{-1}$}
\newcommand{\ergcs}{ergs\,cm$^{-2}$\,s$^{-1}$}
\newcommand{\ergcsa}{ergs\,cm$^{-2}$\,s$^{-1}$\,$\AA^{-1}$}
\newcommand{\cmc}{cm$^{-3}$}
\newcommand{\mic}{$\mu$m}
%
%
\newcommand{\tft}{$\Psi(\tau)$}
\newcommand{\tfvt}{$\Psi(v,\tau)$}
\newcommand{\lvt}{$L(v,t)$}
\newcommand{\lt}{$L(t)$}
\newcommand{\ct}{$C(t)$}
\newcommand{\tlt}{$\tau_{{\scriptscriptstyle LT}}$}
\newcommand{\fvar}{$F_{var}$}
\newcommand{\rmax}{$R_{max}$}
\newcommand{\bdt}{$\Delta t$}
\newcommand{\dff}{$\Delta F/F$}
%
%
\newcommand{\caiik}{\ion{Ca}{2}\,K\,\lb3933}
\newcommand{\cn}{CN\,\lb4200}
\newcommand{\gband}{CH\,Gband\,\lb4301}
\newcommand{\mgi}{\ion{Mg}{1}}
\newcommand{\mgb}{\ion{Mg}{1}+MgH\,\lb5175}
%
%
\newcommand{\brgama}{Br$\gamma$}
\newcommand{\pabeta}{Pa$\beta$}
\newcommand{\palfa}{Pa$\alpha$}
\newcommand{\halfa}{H$\alpha$}
\newcommand{\hbeta}{H$\beta$}
\newcommand{\hgama}{H$\gamma$}
\newcommand{\hdelta}{H$\delta$}
\newcommand{\heps}{H$\epsilon$}
\newcommand{\lya}{Ly$\alpha$}
%
%
\newcommand{\heii}{\ion{He}{2}\,\lb4686}
\newcommand{\hen}{\ion{He}{1}\,\lb5876}
\newcommand{\feii}{\ion{Fe}{2}}
%
%
\newcommand{\nif}{[\ion{N}{1}]\,\lb5199}
\newcommand{\niid}{[\ion{N}{2}]\,\lb6548,6584}
\newcommand{\oii}{[\ion{O}{2}]\,\lb3727}
\newcommand{\oiii}{[\ion{O}{3}]\,\lb5007}
\newcommand{\sii}{[\ion{S}{2}]\,\lb6717,6732}
\newcommand{\neiiia}{[\ion{Ne}{iii}]\,\lb3869}
%
%
\newcommand{\oiiit}{[\ion{O}{3}]\,\lb}
\newcommand{\neiiit}{[\ion{Ne}{3}]\,\lb}
\newcommand{\niit}{[\ion{N}{2}]\,\lb}
%
%
\newcommand{\heuv}{\ion{He}{2}\,\lb1640 + \ion{O}{3}]\,\lb1663}
\newcommand{\civ}{\ion{C}{4}\,\lb1549}
%
%
\newcommand{\ntset}{NGC\,3783}
\newcommand{\ifttn}{IC\,4329A}
\newcommand{\esoofo}{ESO\,141$-$G55}
\newcommand{\aknotz}{Akn\,120}
\newcommand{\frln}{Fairall\,9}
\newcommand{\nfofo}{NGC\,4151}
\newcommand{\pa}{$A$}
\newcommand{\pp}{$p$}
\newcommand{\pco}{$C_o$}
\newcommand{\pin}{$i$}
\newcommand{\ppo}{$\Psi_o$}
\newcommand{\pv}{$V_{sys}$}

\title{HST FOC spectroscopy of the NLR of \nfofo. I. Gas kinematics.
\footnote{Based on observations made with the NASA/ESA {\it Hubble
Space Telescope}, obtained at the Space Telescope Science Institute,
which is operated by the Association of Universities for Research in
Astronomy, Inc., under NASA contract NAS\,5-26555.}}

\author{Cl\'audia Winge\,\altaffilmark{2,3}, David J. Axon\,\altaffilmark{4,5}, 
F.D. Macchetto\,\altaffilmark{4}}

\affil{Space Telescope Science Institute, 3700 San Martin Drive, Baltimore, 
MD21218, USA.}

\author{A. Capetti}

\affil{Osservatorio Astronomico di Torino, Strada Osservatorio 20, 10025 Pino 
Torinese, Torino, Italy.}

\and

\author{A. Marconi}

\affil{Osservatorio Astrofisico di Arcetri, Largo E. Fermi 5, 50125 Firenze, 
Italy.}

\altaffiltext{2}{CNPq Fellowship, Brazil; winge@if.ufrgs.br}

\altaffiltext{3}{Current address: Instituto de F\'{\i}sica,
Universidade Federal do Rio Grande do Sul, Av. Bento Gon\c{c}alves,
9500, C.P. 15051, CEP 91501-950, Porto Alegre, RS, Brazil.}

\altaffiltext{4}{Affiliated to the Space Science Department of ESA}

\altaffiltext{5}{Division of Physical Sciences, University of Hertfordshire, College Lane, Hatfield, Herts, AL10 9AB, U.K}

\begin{abstract}

We present the results from a detailed kinematic analysis
of both ground-based, and {\it Hubble Space Telescope}/Faint Object
Camera long-slit spectroscopy at sub-arcsec spatial resolution, of the
narrow-line region of \nfofo. In agreement with previous work, the
extended emission gas ($R > $4\arcsec) is found to be in normal
rotation in the galactic plane, a behaviour that we were able to trace
even across the nuclear region, where the gas is strongly disturbed by
the interaction with the radio jet, and connects smoothly with the
large scale rotation defined by the neutral gas emission.

The HST data, at 0\arcsec.029 spatial resolution, allow us for the
first time to truly isolate the kinematic behaviour of the individual
clouds in the inner narrow-line region. We find that, underlying
the perturbations introduced by the radio ejecta, the general
velocity field can still be well represented by planar rotation down to a 
radius of \app\ 0\arcsec.5 (30 pc), distance at which the rotation 
curve has its turnover.

The most striking result that emerges from our analysis is that the
galaxy potential derived fitting the rotation curve changes from a
``dark halo'' at the ENLR distances to  dominated by the central mass
concentration in the NLR, with an almost Keplerian fall-off in the
1\arcsec $<$ R $<$ 4\arcsec interval.  The observed velocity of the gas
at 0\arcsec.5 implies a mass of M \app\ \tento{9} \msol\ within the
inner 60 pc.  The presence of a turnover in the rotation curve
indicates that this central mass concentration is extended. The first
measured velocity point (outside the region saturated by the nucleus)
would imply an enclosed mass of \app\ 5 \vzs\ \tento{7} \msol\ within R
\app\ 0\arcsec.15  (10 pc) which represents an upper limit to any
nuclear point mass.

\end{abstract}


\section{Introduction} \label{sec_intro}

The SABab galaxy \nfofo\ hosts one of the most studied active galactic
nuclei (AGN) in the sky, with observations from the $\gamma$-ray to the
MHz range (e.g., Kriss et al.\markcite{gketal95} 1995; Knop et
al.\markcite{rketal96} 1996; Warwick et al.\markcite{rwetal96} 1996;
Ulvestad et al.\markcite{urcw98} 1998).  The AGN spectrum presents a
broad X-ray Fe K$\alpha$\ line (Yaqoob et al.\markcite{yyetal95} 1995),
optical and UV variable broad permitted emission lines (Crenshaw et
al.\markcite{dcetal96} 1996; Kaspi et al.\markcite{sketal96} 1996), as
well as several blue-shifted narrow absorption systems (Weymann et
al.\markcite{wmgh97} 1997), and strong narrow lines originated in a
system of clouds with up to a few kpc extension (P\'erez et
al.\markcite{epetal89} 1989; Yoshida \& Ohtani\markcite{yo93} 1993).
The Extended Narrow Line Region (ENLR) of \nfofo\ shows line
intensities and widths as well as kinematics consistent with
quiescent gas in normal rotation in the galactic disk, illuminated by
the central source (Penston et al.\markcite{mpetal90} 1990; Robinson et
al.\markcite{aretal94} 1994).  Several authors have remarked on the
continuity between the kinematics of the large scale \ion{H}{1} 21 cm
emission and that of the ENLR (Vila-Vilar\'o et
al.\markcite{bv-vetal95} 1995; Asif et al.\markcite{maetal97} 1997), but 
because of the contamination by the Narrow Line Region (NLR; R $<$ 4\arcsec) 
the ground based data have not yielded convincing evidence for normal 
rotation in the circum-nuclear region of NGC 4151 (Schulz\markcite{s:h87} 1987; Mediavilla, Arribas \& Rasilla\markcite{mar92} 1992).

Narrow-band {\it Hubble Space Telescope} (HST) images (Evans et
al.\markcite{ieetal93} 1993; Boksenberg et al.\markcite{abetal95} 1995)
suggested that the complex morphology of the emission clouds and
filaments in the Narrow Line Region (NLR, $R < 4$\arcsec) is shaped not
by the anisotropic character of the central source's radiation but
mainly by the interaction between the hot plasma of the radio jet and
the ambient gas in the disk. In a previous paper (Winge et
al.\markcite{cwetal97} 1997, hereafter Paper I), we presented the initial
results from HST long-slit spectroscopy of the NLR of \nfofo\ at a
spatial resolution of 0\arcsec.029, isolating for the first time the
spectra of individual clouds and demonstrating the influence of the
radio ejecta in both the local kinematic and ionization conditions of
the emission gas. Several very localized sub-systems of both blue and
red-shifted high-velocity knots were detected, and we also observed
off-nuclear continuum emission and marked variations on the emission
line ratios within a few pc. Such evidence indicates that the physical
conditions of the emission gas in individual clouds are strongly
influenced by local parameters, either density fluctuations or shock
ionization, possibly both.

In this paper we present a detailed study of the kinematics of the gas
in both the extended and inner NLR of \nfofo\ from ground-based and HST
data with high spatial resolution that allow us to separate the
underlying velocity field of the emission gas in the NLR from the
effects of the radio jet,  and to probe its connection with the large
scale rotation of the ENLR in the galactic disk. The most striking
evidence for the interaction of the radio jet with the ambient gas and
the presence of strong shocks, observed as high-velocity emission
knots, localized off-nuclear continuum emission, and variations in the
emission-line ratios in scales of a few to tens of parsec, will be
discussed in a forthcoming paper.

For a distance to \nfofo\ of 13.3 Mpc, 0\arcsec.1 corresponds to a
linear scale of 6.4 pc in the plane of the sky; a value of \hbc\ = 75
\kms Mpc$^{-1}$ is assumed throughout this paper.

\section{Observations and Data Reduction} 

\subsection{Ground Based data} \label{subsec_obsgb}

Long-slit spectroscopic observations of NGC4151 were obtained using the
IPCS (Boksenberg\markcite{b:a72} 1972; Boksenberg \&
Burgess\markcite{bb73} 1973) at the f/15 Cassegrain focus of the 2.5m
Isaac Newton Telescope on March 9 to 13, 1985. The wavelength coverage
includes the \oiiit 4959,5007 and \hbeta\ lines, at an
instrumental resolution of 0.75 \AA\ (45 \kms) Full Width Half Maximum
(FWHM). The  spatial resolution was 0\arcsec.63. A total of 17 spectra
were obtained at PA = 48\arcdeg\ and 138\arcdeg\ (parallel and
perpendicular to the ENLR direction, respectively) at several different
offsets from the nucleus. Different slit widths were used, from
0\arcsec.22 to 0\arcsec.65. The slit positions are shown in
Figure~\ref{fig_slitsgb} and the log of the observations is given in
Table~\ref{tab_loggb}.

The individual frames were reduced using the FIGARO image processing
environment (Shortridge\markcite{s:k93} 1993), and then re-binned to a
resolution of 0.24 \AA/pixel (60 \kms).  The continuum spectra were
subtracted after correction for vignetting effects,  and the spatial
distribution of emission line fluxes, central velocity and FWHM
obtained by fitting Gaussian functions to the line profiles using the
LONGSLIT spectral analysis software (Wilkins \& Axon\markcite{wa92}
1992).

\subsection{HST FOC f/48 data} \label{subsec_obsst}

The NLR of \nfofo\ was observed using the HST Faint Object Camera (FOC) 
f/48 long-slit spectrograph on July 3, 1996. The slit, 0\arcsec.063
\vzs\ 13\arcsec.5 in size,  was positioned along PA = 47\arcdeg. The
F305LP filter was used to isolate the first order spectrum which covers
the 3650 -- 5470 \AA\ interval at a 1.58 \AA/pixel resolution. The
spatial scale is 0\arcsec.0287 per pixel and the instrumental PSF is of
the order of 0\arcsec.08. The observational procedure consisted in obtaining  
first an interactive acquisition (IntAq) image in 1024 x 512 zoomed
mode with the f/48 camera through the F220W + F275W filters to allow
for an accurate centering of the object. While the necessary offsets
were calculated, a 1247 seconds, 1024 \vzs\ 512 non-zoomed mode 
spectrum  was taken with the slit at the IntAq position. Six 697 
seconds (non-zoomed) spectra were then obtained stepping across 
the NLR in 0\arcsec.2 intervals, starting 0\arcsec.6 SE of the nucleus. 
The three resulting SE spectra were too faint to be useful. The slit positions, 
derived as described below, are shown in Figure~\ref{fig_slits96}, 
superimposed on an  archival \oiii\ FOC f/96 image of the galaxy and 
listed in Table~\ref{tab_logst}.

The spatial relation between objects in the sky is not preserved in
the uncalibrated data from the cameras since the raw FOC images are
affected by strong geometric distortion.  This distortion comprises two components:
the external or optical, due to the off-axis position of the detector
itself; and the internal, a combination of the distortion caused by the
magnetic focusing of the image intensifiers and the one introduced by
the spectrographic mirror and grating. The internal contribution is by
far the most important, and it is strongly time and format dependent.
We have used both standard IRAF\footnote{IRAF is distributed by the
National Optical Astronomy Observatories, which are operated by the
Association of Universities for Research in Astronomy, Inc., under
cooperative agreement with the National Science Foundation.} procedures
for spectroscopic data reduction as well as specific packages developed
for the FOC data (at stsdas.hst\_calib.foc) to calibrate the data.

Initially, all frames, including those used for subsequent
calibration, were geometrically corrected for the optical plus focusing
induced distortions using the equally spaced grid of reseaux marks
etched onto the first photocatode in the intensifier tube. The observed
position of the reseaux marks were measured in the internal flat-field
frames bracketing the observations and then compared with an equally
spaced artificial grid of suitable size (9 \vzs\ 17 reseaux marks for the 1024
\vzs\ 512 mode) already corrected by the {\it inverse} optical
distortion\footnote{determined from ray-tracing models of the HST and
FOC optics and available within IRAF.}. Each individual transformation
was computed fitting two dimensional Chebyshev polynomials of 6$^{th}$
order in x and y and 5$^{th}$ order in the cross-terms, and applied to
the respective science and calibration frames. The rms uncertainties in
the reseaux positions are 0.12 -- 0.20 pixel for the 512 mode,
depending on the signal to noise (S/N) ratio of the flat-field images.

The  remaining (mirror$+$grating) internal distortion along the
dispersion direction was corrected by tracing the spectra of two stars
observed in the core of the globular cluster 47 Tuc. The stars are
\app\ 130 pixels apart and provide a good correction for most
of the working area of the slit. The rms of the tracing is about 0.2
pixel. The distortion along the spatial direction was obtained in a
similar way, tracing the brightness distribution of the emission lines
of the planetary nebula NGC\,6543, with residuals of 0.10 -- 0.22
pixel.  The two corrections were combined in a single calibration file
and applied simultaneously to the science frames.

Since the FOC spectrograph does not contain an internal reference
source, the NGC\,6543 geometrically-corrected frame was used to obtain
the wavelength calibration. Reference wavelengths were derived from
ground-based observations (P\'erez et al., in preparation), which also
indicate that distortions introduced by the internal velocity field of
the nebula are negligible at the f/48 resolution  (less than 0.5
\AA\ \app\ 0.3 pixel). The two-dimensional spectrum was collapsed along
the slit, and a 6$^{th}$ order Legendre polynomial solution found for
the pixel-to-wavelength transformation, with residuals of \app\ 0.04
\AA. The lines were re-identified in the original frame and a
bi-dimensional wavelength calibration file obtained. The spatially
extended emission lines in the fully calibrated frame of NGC\,6543 are
measured to be within 0.13 \AA\ of their reference values.
 
The procedure above can be summarized as follows:
\begin{enumerate}
\item optical plus main geometric distortions are corrected using the
reseaux marks and left a residual error of 0.12 -- 0.20 pixel.
\item residual (spectrograph) distortion in the dispersion direction 
is corrected tracing the spectra of point sources and  the  error is 
\app\ 0.2 pixel.
\item residual distortion along the slit is corrected tracing the 
emission lines of an extended source and the error is 0.10 -- 0.22 pixel.
\item errors from internal velocity field of the wavelength calibrator
and from the calibration itself are less than 0.5 and 0.13 \AA,
respectively.
\end{enumerate}

The final errors estimated for the \nfofo\ spectra, combining all the
above sources in quadrature (including the errors from the geometric
corrections applied to the calibration files),  are 0\arcsec.016 and 1.1 \AA\ rms in the 
spatial and dispersion directions,
respectively. The final  wavelength resolution is \app\ 310 \kms\ at
\oiii. 

Flux calibration was obtained using the UV standard star LDS749b. The
data frame was geometrically corrected and wavelength calibrated as
above, and the spectrum extracted on a 16-pixel window. This was then
divided by the integration time and by the appropriate segment of the
absolute flux table, rebinned to match the wavelength interval. A
6$^{th}$ order spline3\footnote{The ``order'' of a liner or cubic spline function in the IRAF routines refers to the number of polinomial pieces in the sample region.} was fitted to the resulting counts per flux per second spectrum, averaged in 60 \AA\ intervals, generating a smooth
response function. The vignetting along the spatial direction was
corrected using the model presented in the FOC Handbook (Nota et
al.\markcite{anetal96} 1996), and combined with the above response
function to obtain a two-dimensional sensitivity calibration frame. We
estimate light-losses due to the small size of the slit to be in the
order of 20\% for a point source. The relative flux of lines measured
within a 0\arcsec.3 -- 0\arcsec.6 (10 -- 20 pixels) interval is believed
to be accurate at a 10 -- 15\% level.

The science frames were divided by the exposure time and by the
composite sensitivity frame. The background emission was subtracted by
fitting a spline3 function along both spatial and dispersion directions,
after masking the regions with emission lines and continuum. We opted
to keep the fitting function order as low as possible which provides
the best subtraction over most of the frame, even when this implied an
imperfect result over small areas, where the background was larger
and/or a more rapidly varying function of
 position. An example of the resulting 2D frames is shown in
Figure~\ref{surface_fig}. The plots correspond to a 2\arcsec.6 \vzs\ 71
\AA\ (166 pc \vzs\ 4250 \kms) segment of the \oiii\ emission line
centered on the nuclear continuum on the PA47\_1 spectrum (top) and
0\arcsec.6 SW from the center of the slit on the PA47\_3 spectrum
(bottom).  Multiple velocity systems and a complex cloud structure are
seen, and some features can be easily identified in the spectra
show in Figure~\ref{fig_gaussfit}, like the broad plume 0\arcsec.37 SW
of the nucleus in the top image or the double-peaked feature and
high-velocity cloud 0\arcsec.2 and 1\arcsec\ SW, respectively, in the
0\arcsec.41 NW image.

To accurately determine the slit positions, we compared the
\oiii\ luminosity profile of each spectrum with that derived from an
archive FOC f/96 image taken with the F501N filter. The observed
spectra were convolved with the transmission curve of the filter and
the data collapsed along the dispersion direction. The f/96 image was
rebinned to match the f/48 spatial resolution and the spectra light
profiles compared with the sum of two successive lines stepping across
the image at PA = 47\arcdeg. The final agreement is good within a few
percent and the final uncertainty in the position corresponds to half
the size of the slit (or one line in the image), 0\arcsec.03.
  
We also retrieved from the HST Archive the \nfofo\ spectra obtained on
July 11, 1995 as an engineering test of the f/48 detector. The slit was
positioned at PA = 40\arcdeg\ and the spectra were taken using the 1024
\vzs\ 256 mode.  Data reduction was complicated by the lack of an
equivalent flat-field frame (the ones interspersed with the
observations were taken in the 1024 \vzs\ 512 zoomed mode), and
contemporaneous wavelength calibration observations, resulting in larger
uncertainties in both spatial and spectral scales. Due to the higher
background features located in the upper and middle part of the frames,
as well as the presence of a serious ``blemish'' in the detector crossing 
over part of the \oiiit 4959,5007 emission lines, we have chosen to
optimize the reduction for the \oii\ region, resulting in final
uncertainties of 0\arcsec.02 and 1.24 \AA\ in the spatial and dispersion
directions, respectively. We note, however, that several emission and
kinematic features observed in the 1996 [\ion{O}{3}] data can also be 
identified in the 1995 frames (see Section~\ref{sec_res}). Five data
sets contain enough signal to be useful, and their positions, derived
in a similar way to the 1996 data, are listed in Table~\ref{tab_logst}
and shown in Figure~\ref{fig_slits95}.

\section{Results} \label{sec_res}

\subsection{Ground-based Data} \label{subsec_resgb}

Figure~\ref{fig_gaussgb} shows a selection of the \oiii\ line profiles
observed along the PA48 position at various offsets from the nucleus.
In the inner 5\arcsec\ (the NLR), up to 3 Gaussian components were
needed to fit the line profiles, while the extended emission was well
represented by a single Gaussian with the instrumental resolution
(\app\ 45 \kms). The transition between the ENLR and the inner NLR at
around 4\arcsec\ is particularly instructive.  Here, the extended
narrow component is joined by a second blue-shifted component of \app\ 400
\kms\ FWHM. Interior to this radius the line profiles are always broad
and double.

Figure~\ref{fig_vobsgb} presents the result of the profile analysis for
the PA48 position. It can be seen that the velocity structure of the
extended emission (represented as stars) is similar to that of a
typical galactic rotation curve, indicating that this emission
originates from gas in the galactic plane. The line width of $<$ 45
\kms\ is also consistent with the velocity dispersion of gas in the
disks of normal spiral galaxies (van der Kruit \&
Shostak\markcite{ks84} 1984). This component has a ratio
[\ion{O}{3}]/\hbeta\ \app\ 9, indicating it is photoionized by the AGN
continuum. The general impression is that the FWHM of this primary
component increases steadily as one approaches the nucleus while the
radial velocity connects almost seamlessly onto the velocity field of
the extended narrow component, a point we shall re-emphasize when we
discuss the FOC f/48 results. This progression is well illustrated by
the line profiles between 1\arcsec.9 and 3\arcsec.8 SW
(Figure~\ref{fig_gaussgb}c and \ref{fig_gaussgb}d). The second
component (open squares) is much broader, FWHM \app\ 600 \kms, is
present only in the inner 5\arcsec\ of the emission region, and is
systematically blueshifted with respect to the primary. The
[\ion{O}{3}]/\hbeta\ ratio is \app\ 7. At first sight, the physical
reality of the third, narrower, component (FWHM \app\ 100 \kms), which
is  only found very close to the nucleus, with high velocity shifts
with respect to the main rotation curve, may be thought as
questionable.  However, as we shall see, it corresponds to structures
identified in the high-spatial resolution FOC f/48 data, closely
associated with the radio jet.

\subsection{HST FOC f/48 Data} \label{subsec_resst}

To study the kinematics of the gas in the HST spectra the brightest
emission lines, \oii\ and \oiiit 4959,5007, were extracted in spatial
windows varying from 2 to 20 pixels along the slit, and their profiles
fitted using 1 to 3 Gaussian components. The two [\ion{O}{3}] lines
were fitted simultaneously, constraining the physical parameters of the
corresponding components of each line to their theoretical values (3:1
intensity ratio, 48 \AA\ separation, and same FWHM).
Figure~\ref{fig_gaussfit} shows some examples of the fits for the 1996
[\ion{O}{3}] region. The spectra are identified by their distance to
the center of the slit, defined as the line passing through the
nucleus perpendicular to the slit direction, with negative values
running towards the SW.  All regions plotted are located within
1\arcsec.5 from the active nucleus and the line profile representations
vary from a single component, with very little broadening (second
spectrum at position PA47\_3) to combinations of very broad or multiple
Gaussians.

Figure~\ref{fig_gauss95} shows two regions of the \oiiit 4959,5007
emission in the 1995 0\arcsec.14  and 0\arcsec.52 NW (PA40\_3 and
PA40\_5) frames, which correspond to the same kinematical features
already remarked in Figure~\ref{surface_fig}, the nuclear plume and
the double peaked profile. Even with the presence of the blemish
redward of \oiii, the similarity with the nearby 1996 profiles is
evident.

The results of the decomposition are shown in Figures~\ref{fig_vobso3}a
to \ref{fig_vobso3}d for the 1996 \oiii\ data, where the top panels
show the FWHM of the different components superimposed over the
brightness profiles of the radio emission (from the VLA$+$Merlin 5 GHz
radio map of Pedlar et al.\markcite{apetal93} 1993). In these figures,
the data are plotted as a function of position along the slit. The
resulting components  naturally separate into three groups: a narrow
(FWHM $\lesssim$ 700 \kms), closest to the systemic velocity and
usually the brightest component in the fit (filled triangles); a broad
(FWHM $\gtrsim$ 700 \kms) base that appears in the highest S/N spectra
and is distributed around the radio knots (stars); and the
occasional narrow secondary component, which can be split from the main
emission by as much as \mm\,1000 \kms\ (open circles).  The gray region
is the instrumental FWHM. The seven points on the PA47\_1 (nuclear)
position plotted as open triangles will be discussed later (see
Section~\ref{subsec_fitst1}).

As discussed in Paper I, there is an association between the optical
and the radio emission, in the sense that the brightest emission-line
filaments surround the radio knots, as expected in a scenario where the
plasma of the radio jet is clearing a channel in the surrounding medium
and enhancing the line emission along its edges by compression of the
ambient gas (Taylor, Dyson \& Axon\markcite{tda92} 1992; Steffen et al.
1997a\markcite{wsetal97a} ,b\markcite{wsetal97b} ).
Figures~\ref{fig_vobso3}a -- \ref{fig_vobso3}d present the kinematic
expression of this association, with broad bases and/or high velocity
components closely associated with the radio emission, while the
``main'' narrow component follows a more ordered pattern,
strikingly resembling that of a disk rotation curve.

The lower resolution  (instrumental FWHM \app\ 450 \kms) and lower S/N
in the \oii\ region did not allow us to isolate as many components as for
\oiii. Nevertheless, the overall behaviour of the velocity field, shown in
Figures~\ref{fig_vobso2} and \ref{fig_vobso2a} for the 1996 and 1995
data, respectively, is the same as for the \oiii\ emission, with the
narrow component tracing rotation in the disk, while the broad and
secondary components mark the interaction of the radio plasma with the
ambient gas.

Based on our observations, we propose that the emission gas in the
inner 5\arcsec\ of \nfofo\ is, to the first order, produced by the
central source's photoionization of the ambient gas located in the disk
of the galaxy, and therefore, its kinematic behaviour is mainly
planar rotation, determined by the dominant potential in the nuclear
region (either a very concentrated but extended mass distribution or a central Massive 
Dark Object). The ionizing photons escape along a broad cone which grazes the
disk, as suggested by Pedlar et al.\markcite{phau92} (1992), and
Robinson et al.\markcite{aretal94} (1994). From our data we can see
that the systematic outflow previously remarked on the literature (e.g., 
Schulz\markcite{s:h90} 1990) as
dominating the kinematics of the emission gas in the NLR of \nfofo\ can be 
understood as an effect of the lack of spatial resolution of the ground based data. When the spectra of the individual clouds are obtained, it is possible
to decouple the gas that is in general rotation in  the disk under the
gravitational influence of the central mass concentration from that
which is being entrained and swept along by the radio plasma, as it
plunges through the galactic disk.  This scenario has already been
suggested by Vila-Vilar\'o et al.\markcite{bv-vetal95}  (1995), based
on a higher (0.34 arcsec/pixel and \app\ 1\arcsec\ seeing) spatial resolution ground-based
spectrum oriented along PA = 51\arcdeg, where they found that the
emission within 3\arcsec\ from the active nucleus can be decomposed
into one blue-shifted component located mainly SW of the nucleus, and
a second system that they describe as appearing ``to link the
blue-shifted ENLR SW of the nucleus with the red-shifted emission to
the NE. It is conceivable that it represents a continuation of the ENLR
velocity field and hence traces the galactic rotation curve within the
NLR.'' Mediavilla et al.\markcite{mar92} (1992) also remarked on the
existence of a gradual connection between the kinematic and physical
properties of the galactic environment and the (ground-based) unresolved
inner NLR.

We also note that if the scenario of an outflow along the edges of the
ionization cone is invoked, it is necessary for the line of sight be
located outside the cone or systematic blue-shifted components would be
projected on both SW and NE sides, which is not observed. A narrow
cone, however, implies a \app\ 25\arcdeg\ misalignment between the
outer ENLR and the direction of the radio jet, and therefore to a
geometry that is distinct from the ``standard'' unified model. Pedlar
et al.\markcite{apetal93} (1993) first remarked on the presence of
small, systematic changes in the jet orientation, since the position
angle of the jet in the inner 3\arcsec\ SW oscillates between
254\arcdeg\ and 263\arcdeg\ (74\arcdeg\ to 83\arcdeg), with an
approximate mirror symmetry to the NE side. The abrupt change of
55\arcdeg\ in  the direction of the radio emission in the milliarcsec
scale as seen in the VLBA radio maps of Ulvestad et
al.\markcite{urcw98} (1998) suggests that  effects such as precession or
warping are very likely present in the collimating structure around the
central source (Pringle\markcite{p:j97} 1997), but the fact that the
arcsecond-scale radio jet is aligned within 10\arcdeg\ indicates that
these effects would be transient and their impact in the alignment and
morphology of the NLR/ENLR/radio emission are difficult to determine.

The presence of the variable blue-shifted optical and UV absorption
lines is clear evidence that a gas outflow {\it is} present in the
nuclear region of \nfofo. The high resolution GHRS spectra presented by
Weymann et al.\markcite{wmgh97} (1997) show the existence of several
distinct systems with outflow velocities from 300 to  1600 \kms\ with
respect to the nucleus, but the material responsible for these features
is very likely to be located inside 1 pc of the active nucleus and the
actual geometry and dynamics of the clouds is not known (Espey et
al.\markcite{bretal98} 1998). Although tempting, to associate the bulk
kinematics of the emission-line gas of the NLR on scales of several to
hundreds of parsecs with the nuclear outflowing gas is not consistent
with our data, since we observe very localized clouds with
both blue- {\it and} red-shifted emission within the velocity range 
quoted for the outflows above, while the bulk of the gas can be well
described by a planar rotation model.

\section{Modeling the Rotation Curve} 

To study the gas velocity field, we concentrate on  the narrowest,
closer to the systemic velocity, and most frequently the brightest,
Gaussian component (the filled triangles in Figures~\ref{fig_vobso3} to
\ref{fig_vobso2a}) as representative of  the ``main'' NLR emission. The
same argument was used to define the ground-based rotation curve (see
Section~\ref{subsec_resgb}). In the ENLR, the general behaviour of the
velocity field  connects smoothly with that of the neutral gas in the
kiloparsec scale (e.g.,  Vila-Vilar\'o et al.\markcite{bv-vetal95}
1995). Here, we proceed on the assumption that the gas represented by
the ``main'' component of the velocity field we isolated in the HST data is, to a first 
approximation, also
participating in the general rotation of the gas in the galactic disk,
since it also connects reasonably well with the extended emission, as
can be seen in Figure~\ref{fig_pa48all_fit08}.

\subsection{The Model} \label{subsec_mod}

We have used the Bertola et al.\markcite{fbetal91} (1991) analytic 
expression, which provides a simple parametric representation for 
particles (gas or stars) on circular orbits in a plane, in the form:

\begin{equation}
V_c(r) = V_{sys} +  \frac{Ar}{(r^2 + C_o^2)^{p/2}}
\label{eq:rot}
\end{equation}

\noindent
where \pv\ is the systemic velocity, $r$ is the radius in the
plane of the disk and \pa, \pco, and \pp\ are parameters that define
the amplitude and shape of the curve. If $v(R,\Psi)$ is the
radial velocity at a position $(R,\Psi)$ in the plane of the sky, where
$R$ is the projected radial distance from the nucleus and $\Psi$ its
corresponding position angle, we have:

\begin{equation}
v_{mod}(R,\Psi)  =  V_{sys} + \frac{A\,R\,\cos(\Psi - \Psi_o)\,
\sin i\,\cos^p i}
{\{R^2\eta + C_o^2\,cos^2 i\}^{p/2}}
\label{eq:rotobs}
\end{equation}

\noindent
where 
\begin{displaymath}
\eta \equiv [sin^2 (\Psi - \Psi_o) +  cos^2 i\,cos^2(\Psi - \Psi_o)] 
\end{displaymath}

\noindent
where \pin\ is the inclination of the disk ($i=0$\arcdeg\ for a face-on disk) 
and \ppo\ the position angle of the line of nodes. 

\subsection{The Fitting} \label{subsec_fit}

We used a Levenberg-Marquardt non-linear least-squares algorithm to fit
the above model. The various parameters are determined simultaneously
by minimizing the residuals $\Delta v = (v_{obs} - v_{mod})$, with
$v_{mod}(R,\Psi;A,C_o,p,i)$ and $v_{obs}(R,\Psi)$ being the model and
observed radial velocities at the position $(R,\Psi)$ in the plane of
the sky, respectively.

The rotation curve expressed in Equation \ref{eq:rotobs}, while
giving  a simple representation of the gas kinematics, has a few
shortcomings from the point of view of a minimization procedure. The two
projection angles, \pin\ and \ppo\ can be independently determined only 
if data are available for more than one position angle in the galaxy. 
In a single run of the programme, they are also strongly dependent on the initial values provided to the algorithm.
 
The parameters \pa\ and \pin\ are strongly coupled when determining the
amplitude of the rotation curve, so equally acceptable fits can be
obtained with  a large \pa\ and a small \pin\ or vice-versa. This
situation is more acute for \pp\ \app 1, and the two parameters tend to
decouple for larger values of the exponent. The parameter \pco\ is
determined mainly by the steepness of the inner part of the rotation
curve, implying that the use of off-nuclear slit positions will tend to
push its value to larger radii simply by the absence of data
corresponding to smaller values of $R$. On the other hand, the exponent
\pp\ is determined by the outer parts of the rotation curve, and its
value is expected to be between 1 for a ``dark halo'' potential, and
1.5 for a system with finite mass contained within the ``turn-over'' radius. Evidently, data points for larger values of $R$ will tend to reflect the potential generated by the mass distribution on larger scales.

To test the algorithm, we used the original data (with added errors of 5-10\%) from Bertola et al.\markcite{fbetal91} (1991) on NGC\,5077 and followed the same
procedure described in their paper. Our resulting model agrees with
theirs within an rms of 8 \kms, with the final parameters differing by
less than 5\%. Also, applying the algorithm to an artificial rotation curve with spatial sampling and added errors similar to those of the \nfofo\ data retrieves the original parameter set to better than 1\%.

\subsubsection{The Ground Based data set} \label{subsec_fitg} 

To derive the projection angles of the galaxy's disk, we used the
ground based data listed on Table~\ref{tab_loggb}, and excluded from
the fit the points within 5\arcsec\ from the active nucleus since the
spatial resolution does not allow us to separate the ``main''
rotating component from the highly disturbed gas interacting with the
radio jet in the NLR.

Initially, the data from the three slit positions at PA =
48\arcdeg\ were  fitted with \pa, \pp, \pco, \pin,  \ppo, and
\pv\ as free parameters, with \pv\ having the same value for all data
points. Then, to take into account any residual zero point velocity offsets 
between the data sets, each position angle was separately fitted allowing
\pv\ to vary, but keeping the remaining parameters fixed to the values
obtained before. The resulting systemic velocity for each position was
subtracted from the observed values and the whole data set refitted
with now five free parameters. Once convergence was achieved, the new
model was used to obtain the value of \pv\ at each slit position, and
the procedure repeated. The model was found to be stable at the third
iteration, and the velocity offsets between data sets were smaller 
than 10 \kms.

Finally, the effect of the initial guess on the parameters was explored
running the program for a wide interval of Monte Carlo search for each
parameter, while the others were given their best fit values as the
initial guesses.  The resulting  ranges in the best-fit parameters are
shown in Table~\ref{tab_modgb}. They tend to cluster in two families
characterized by different values in the position angle of the line of
nodes (\ppo). Since the least-squares minimization is unable to select
between them, the two average models for these families are also listed
as Models A and B  and plotted in Figure~\ref{fig_fitgb}.
There we see that our data alone is not enough to allow us to
discard either of these models, so we choose the value of \ppo\ as
33\arcdeg.9, which is closer to the results quoted in the literature
(\ppo\ $=$ 29\arcdeg.1 \mm\ 2.4 at R $=$ 10\arcsec\ from \ion{H}{1} 21 cm
observations -- Pedlar et al.\markcite{phau92} (1992);
\ppo\ \app\ 34\arcdeg -- 41\arcdeg\ at R $=$ 2\arcsec.5 from
\halfa$+$[\ion{N}{2}] measurements -- Mediavilla \&
Arribas\markcite{ma95} (1995)). We stress that the final model obtained
from the HST data is essentially insensitive to the exact choice of
value within this narrow range, and therefore for simplicity, in the
rest of this paper we  present only the analysis carried out using
Model A.

Other than that, the final solution is stable in all 
parameters: the inclination of
the disk, \pin\ = 21\arcdeg, agrees very well with other photometric
and kinematic determinations (Simkin\markcite{s:s75} 1975; Bosma et
al.\markcite{bel77} 1977; Pedlar et al. \markcite{phau92} 1992); the
rotation curve derived from the emission gas in the ENLR is essentially
flat for R $\gtrsim$ 15\arcsec (\app\ 1 kpc), as observed from
\ion{H}{1}\ and optical data (Pedlar et al.\markcite{phau92} 1992;
Robinson et al.\markcite{aretal94} 1994; Vila-Vilar\'o et
al.\markcite{bv-vetal95} 1995; Asif et al.\markcite{maetal97} 1997);
the amplitude \pa\ is typical of the values observed for normal
galaxies with similar absolute magnitude and Hubble type (Rubin et
al.\markcite{rbft85} 1985). On the other hand, the parameter
\pco\ \app\ 440 pc, which for \pp\ = 1 is the radius at which the
velocity reaches 70\% of its maximum value, is smaller than the typical
value of $\geq$\ 1.0 kpc, indicating a large central mass
concentration.

Using the Model A above to obtain the predicted rotation velocity at
the positions sampled by the PA = 138\arcdeg\ data, we found that the
data scatters around the model without much evidence of an ordered
velocity field.  This is not surprising for the three inner positions
(2\arcsec.5 NE, Nucleus, 2\arcsec.5 SW) where perturbations can be
induced by the radio jet plunging through the ambient gas. For the
outermost positions, the observed points present much shallower
gradients than expected from the model, with values smaller by 15 -- 50
\kms. The presence of turbulence on the velocity field across the outer
knots has been already noted by other authors, and explained as effects
of the expansion of the ionization front that produces the line
emission (Asif et al.\markcite{maetal97} 1997) or of cloud-cloud
collisions where the gas streaming in from the leading edge of the bar
meets the one trapped in the inner Lindblad resonance orbits (Robinson
et al.\markcite{aretal94} 1994; Vila-Vilar\'o et
al.\markcite{bv-vetal95} 1995). 

Therefore, in agreement with previous works, we conclude that the
kinematics of the ionized gas in the ENLR  of \nfofo\ beyond a distance
of about 0.5 kpc is dominated by the general rotation of the galactic
disk, with significant perturbations present in the emission-line knots
due to a non-circular or non-planar velocity component associated with
gas turbulence, effects of the ionization front and/or of the presence
of the galactic bar.

Under the hypothesis that the emission  in the inner 5\arcsec\ of the
NLR of \nfofo\ is produced by gas rotating in the disk of the galaxy,
ionized mainly by the central source and kinematically {\it disturbed}
by the interaction with the radio jet, we now assume that the
projection of the velocity field in the plane of the sky is determined
by the same geometrical parameters (\pin\ and \ppo) as obtained from
the ENLR rotation curve, and use the same procedure as above to fit 
the FOC f/48 data.

Notice that we do not expect the other parameters of the fit to be the
same or even similar to those obtained for the large scale rotation,
since the high spatial resolution of the HST data is sampling the gas
whose behaviour is governed by the very inner part of the gravitational
potential well, while the ENLR gas reacts to the mass distribution on
kpc scales.

\subsubsection{The 1996 FOC f/48 [\ion{O}{3}] data set} \label{subsec_fitst1}
 
As before, the full data set was first fitted with \pa, \pp, \pco, and
\pv\ as free parameters, while \pin\ and \ppo\ were kept fixed at
21\arcdeg\ and 33\arcdeg.9, respectively, and with \pv\ having the same
value for all data points, then each slit position was separately
fitted with the resulting model but allowing  \pv\ to vary. As a double
check of the result, we also obtained a fit for each slit position
allowing all four parameters to vary, and found that the values of
\pv\ for the individual ``best'' and  constrained fits agree well
within the errors.  As explained in Macchetto et al.\markcite{dmetal97}
(1997), the repositioning of the FOC spectrographic mirror, which moves
between flat-field and source exposures, can cause a shift in the zero
point of the wavelength scale. If one of the individual data sets is
shifted up or down relative to the others, the simultaneous fit to all
data will result in a weighted value for \pv, and a larger value for
\pa\ than what would be expected from a uniform wavelength scale. We
found a \app\ 10 \kms\ shift between the systemic velocity for position
PA47\_1 and both  PA47\_2 and PA47\_4, obtained in the same orbit, and
a \app\ 50 \kms\ shift between them and PA47\_3, the IntAq spectrum.

The resulting systemic velocity for each slit position was then
subtracted from the observed values and the whole data set refitted
with \pa, \pp, and \pco\ as free parameters. Finally, the effect of the
initial guess on the parameters was explored as for the ground based
data, and the resulting intervals in the parameters are listed in
Table~\ref{tab_modst}, together with the average model (``fit'')  and
the one obtained using only the PA47\_1 spectra (``nucleus'') which, as
discussed in Section~\ref{subsec_fit}, would be more sensitive to the
actual value of \pco.  These two models are plotted in
Figure~\ref{fig_fitst1} as full and dotted lines, respectively. Notice
that the points represented as open triangles in
Figure~\ref{fig_vobso3}a had been excluded from the final fit. Their
projected position corresponds to where the radio jet crosses the slit
and therefore we suspect their high velocity can be due to jet
entrainment. If included in the fit, their effect is to increase the
total amplitude \pa\ of the curve, with little change in the other
parameters. The most striking result that emerges from our analysis is
that the exponent \pp\ of Eq.~\ref{eq:rot} changes from 1 to 1.5, indicating
that the potential goes from ``dark halo'' at the ENLR distances to
dominated by the central mass concentration in the interval 1\arcsec 
$<$ R $<$ 4\arcsec, with the gas rotation curve presenting an almost 
Keplerian fall-off in this range of radii.

\subsubsection{The 1995 and 1996 FOC f/48 [\ion{O}{2}] data sets} 
\label{subsec_fitst2}

The same procedure as for the 1996 [\ion{O}{3}] data set was repeated
using the rotation curves obtained from the 1995 and 1996
\oii\ emission-line measurements. For the 1995 data only the two
innermost positions (PA40\_2 and PA40\_3) were used. The three data
sets give similar results, shown in Table~\ref{tab_modst} and plotted
in Figures~\ref{fig_fitst2a} and  \ref{fig_fitst2b} for the \oii\ 1996 and
1995 data, respectively. The NE side of the 0\arcsec.57 SE
rotation curve is systematically blueshifted with respect to the model
in the inner 1\arcsec.2 from the nucleus. This effect can be associated
with the expansion of the radio component C5 (see Figure 2 of Paper 1),
which is localized just NW of the emission sampled by the slit. The low
S/N of the 1995 \oii\  data does not allow us to completely separate
the disturbed components, but one high velocity component can be seen
at the edge of this region in Figure~\ref{fig_vobso2a}.

\subsection{The HST STIS slitless data} \label{subsec_fitout}

In March 1997, slitless CCD spectra of the NLR of \nfofo\ were taken
with the Space Telescope Imaging Spectrometer (STIS) newly installed on
board HST. A description of the data is given in Hutchings et
al.\markcite{jbhetal98} (1998).  Although the \oiii\ spectral image has
comparable spatial resolution to our data, it suffers from considerable confusion in the inner 2
- 3\arcsec, where the complex velocity systems lead to an ambiguity between
spatial and velocity information. The data are, nevertheless,
worthwhile for our purpose since they contain velocity information
extending to larger distances from the nucleus and covering a wider
area of the NLR than our spectra. We have used the data given in
Table~1 of the Hutchings et al.  paper, and  the same archival WFPC2
\oiii\ image to derive the position relative to the nucleus of the
individual clouds identified in their work. We stress that the position
ascribed by us to the clouds is uncertain by at least 2 WFPC2 pixels
(0\arcsec.1). The high velocity systems detected in the STIS data were
excluded and the remaining data within 6\arcsec\ from the nucleus was
fitted using the same procedure as in the previous Sections, with
\pv\ set to zero (the velocities in the STIS data are given relative to
the nuclear \oiii\ emission).

The resulting set of parameters is essentially identical to the one
obtained for the fit of the FOC f/48 \oiii\ nuclear position alone
(model \oiii\ Nuc. in Table~\ref{tab_modst}): \pa\ = 998 \kms; \pp\ =
1.498; \pco\ = 0\arcsec.75.  The bidimensional representation of these
two fits is presented in Figure~\ref{fig_spiderd}, where the STIS data
and corresponding model, and the \oiii\ 1996 FOC f/48 data and the
Nucleus model of Table~\ref{tab_modst}, are shown on the top and bottom
panels, respectively. The velocity contours are 0 to 240 \kms\ in steps
of 40 \kms\ to the NE and the negative equivalents to the SW.  The
residuals, defined as $V_{res} = V_{rot} - V_{model}$, are
negative  (the model overestimates the observed velocity) on the left,
and positive (the model underestimates the observations) on the right
panels. There is no preferential distribution of positive and negative
residuals as will be expected if the velocity field of the gas was
partly due to a large scale radial component such as a bulk outflow. On
the other hand, the largest residuals tend to concentrate around the
position of the brightest radio knots, indicating that the expansion of
the hot plasma introduces a significant non-planar component to the
motion of the ambient gas.

\section{Discussion}  \label{sec_dis}

\subsection{The rotation curve} \label{subsec_rot}

Our INT data confirmed the general result from both \ion{H}{1} and
other ground based observations that the gas at large scales is in
planar rotation, dominated by the galactic potential. As we approach the
nucleus, it is possible to see in both the INT and HST data the
perturbations introduced by the interaction of the radio jet with the
ambient gas superimposed on the rotation component, which however,
still dominates the bulk velocity field.

The combined effect of higher spatial resolution and profile
decomposition is apparent when we compare our data with the rotation
curve of Robinson et al.\markcite{aretal94} (1994), obtained along PA =
51\arcdeg. The INT curve has a smaller total amplitude in the inner
5\arcsec, 180 \kms\ instead of \app\ 350 \kms. Furthermore, the asymmetry 
remarked on previously in the ground-based data, where the SW peak was 
blue-shifted by 200 \kms, and the NE red-shifted by 50 \kms\ with 
respect to the ENLR, disappears. The line profiles shown in Figure~\ref{fig_gaussgb} show
that such asymmetry is created by the mixing of the high velocity
clouds plus the disk component. The behaviour of the line FWHM also
indicates that when crossing the NLR boundary, the single-component
profile  jumps from 420 \kms\ to a little over 600 \kms\ (see
Figure~16b of Robinson et al. 1994), while  our ``main component'' is
always below 300 \kms.  This value is also the FWHM for the rotational 
component in the FOC f/48 data.

One direct consequence of the order of magnitude improvement in spatial
resolution provided by the FOC f/48 data is that the double-peaked,
blue-shifted lines previously observed 3 -- 4\arcsec\ SW of the nucleus
are not evidence of a systematic bulk outflow but rather a complex
profile created by the gas in the disk plus the disturbed component
entrained by the radio jet. The presence of both blue- and red-shifted
high velocity components in the HST spectrum of individual clouds
reinforces the argument that the interaction of the jet with the ISM of
the host galaxy associated with the collimated radiation from the
central source are key factors in determining the morphology of the
NLR,  but not its bulk kinematics. Using the simple assumption that the
gas in the NLR is in the same plane as the outer ENLR, we have been
able to obtain a good parametrization for the rotation curve within
4\arcsec\ of the central source. Our results indicate that the
kinematics of the  NLR gas  in the interval 30 $\lesssim$\ R
$\lesssim$\ 250 pc is best represented by a thin disk in rotation
around the mass distribution contained within the turn-over radius
(\app\ 0.5\arcsec). The observed velocity of the gas at this radius,
taken as purely Keplerian, would imply a mass of M \app\ \tento{9}
\msol\ within the inner 60 pc. 

Although our nuclear spectrum was saturated in the inner 0\arcsec.3 and
the dispersion of the data points within the turn-over radius is rather
large, the velocities measured there are {\it smaller} than further
out, an effect that cannot be ascribed to the spatial PSF, and that
indicates that the \tento{9}\ \msol\ above is not in a point mass but
in an extended distribution. If we assume spherical symmetry for
such mass distribution, our first measured velocity points to the NE
and SW of the nucleus would imply an enclosed mass of \app\ 5
\vzs\ \tento{8} \msol\ within R \app\ 0\arcsec.34 (20 pc; $v_{obs}$
\app\ 100\kms) or \app\ 5 \vzs\ \tento{7}  \msol\ within R
\app\ 0\arcsec.15  (10 pc; $v_{obs}$\app\ 40 \kms), respectively.  This
value would be an upper limit to any point mass that could be present
there.  Further observations with higher (or as high) spatial
resolution but with better sampling in the inner region of the rotation
curve and higher spectral resolution should be able to further
constrain the nature of the mass distribution.

The presence of a large central mass concentration was already
indicated by the small value of the \pco\ parameter when fitting the
kinematics the ENLR gas (see section \ref{subsec_fitg}), and the change
of behaviour of the resulting rotation curve (from flat outside to
Keplerian-like in the inner regions) indicates that it dominates the
kinematics of the NLR. This is confirmed by a quick analysis of the
brightness profile from the archival HST/WFPC2 images. Assuming a
mass-to-light ratio of 2, we find that the stars contribute at most \app\ 2 \vzs\ \tento{8}\ \msol\ to the mass between
0\arcsec.5 and 4\arcsec.  Comparing with a central mass of 5
\vzs\ \tento{8} -- \tento{9}\ \msol\ within a 0\arcsec.5 radius and
considering that the circular velocity scales on the square root of the
mass, the presence of an extended (bulge) component would cause the gas
in the NLR to deviate from a pure Keplerian rotation curve by
$\lesssim$\ 10 -- 20\%. This value is well within the observational
scatter of the data even bearing in mind the considerable
uncertainties introduced by the effect of the central source to the
brightness profile. Thus even though the mass inside the turnover
radius is not a true point source, it is clear why the models yielded
such a closely Keplerian value of \pp\ = 1.5 in our analysis.

The wide-angle ionizing cone model of Pedlar et al.\markcite{apetal93}
(1993) implies that the common collimation axis of the ionizing cone and
the radio jets is not perpendicular to the galactic disk. For
\pin\ \app\ 21\arcdeg, and a bicone opening angle of
\app\ 130\arcdeg\ (see Figure 8 of Boksenberg et al.\markcite{abetal95}
1995), the angle between the collimation axis and the galaxy rotation
axis is \app\ 25\arcdeg, and the one between the collimation axis and
the line of sight, \app\ 40\arcdeg. Our model for the kinematics of the
gas in the inner NLR indicates that it is still rotating in the plane
of the galaxy even at distances of a few tens of parsecs from the
nucleus, and therefore is not directly related to the symmetry axis of
the AGN itself.  Such lack of common orientation is also indicated by
the radio observations of Weymann et al.\markcite{wmgh97}  (1997) at
subparsec scales, and by the Fe K$\alpha$\ profile presented by Yaqoob
et al.\markcite{yyetal95} (1995). If it is assumed that the Fe
K$\alpha$\ line is produced in the accretion disk,  the models indicate
that the structure within \tento{3}\ $r_g$, the gravitational radius of
the black hole, is essentially face-on. A similar situation is seen in
the recent HST infrared observations of Centaurus A (Schreier et
al.\markcite{esetal98} 1998), where the gas disk structure at \app\ 20
pc scales is oriented along the major axis of the bulge, an indication
that its geometry is set by the galaxy gravitational potential rather
than by the symmetry of the AGN and its jet.

\subsection{Evidence for rotation in other objects}

Two other AGN, NGC\,1068 and Mrk\,3, have been spectroscopically
studied with HST with enough detail and spatial resolution to carry
out an analysis similar to that in this paper. While in the first
object we also observe strong but localized perturbations induced by
the interaction of the radio jet with the ambient gas superimposed in
a more general pattern characteristic of ordered rotation (Axon et
al.\markcite{daetal98} 1997), in Mrk\,3 the gas motions in the region
co-spatial with the radio jet are clearly dominated by the expansion
of the cocoon of hot gas shocked and heated by the radio ejecta
(Capetti et al.\markcite{acetal99} 1999).  Such observations are in
agreement with Nelson \& Whittle \markcite{nw96} (1996) results, where it was argued that the correlations between the stellar velocity dispersion and the
[\ion{O}{3}] profile in a large sample of ground-based observations of
Seyferts indicate that the motions in the NLR are predominantly
gravitational in nature, with objects with linear radio sources
presenting broader [\ion{O}{3}] lines.

Evidence for an underlying rotational component is also found in
ground-based studies of several other objects, like NGC\,1365 (Hjelm \&
Lindblad\markcite{hl96} 1996), NGC\,3516 (Mulchaey et
al.\markcite{jmetal92}  1992; Arribas et al.\markcite{saetal97} 1997),
NGC\,2992 (M\'arquez et al.\markcite{imetal98} 1998; Allen et al.\markcite{maetal98} 1998),  with a general
trend for the low-ionization gas to be a better tracer of the disk
component, while the high-ionization gas presents more deviant
behaviour, usually associated with outflow. 

We remark that all these studies used the centroid of the emission
lines, rather than the individual components and, until now, the
observations did not have enough spatial resolution to separate the
individual clouds, and show whether the often noted double-peaked
profile is seen everywhere - implying that the outflow is actually a
wind; or localized - as expected when the gas is entrained by the radio
jet. The observations are also naturally biased toward the brighter
emission line knots, which are more likely to be disturbed, either by
the interaction with the radio jets, tidal effects or even the central
source radiation. Obtaining good S/N data of the central regions of
other AGN with  high spatial resolution and even moderate spectral
resolution can prove to be a useful technique for detecting the undisturbed
gas and probing the nature of the central potential.  Since stellar
dynamical methods are rendered impotent, as the absorption lines in
Seyfert spectra are inevitably filled in by the featureless continuum,
this approach may provide the only means of directly determining the
central object (black hole) mass in currently active nuclei.

A potentially interesting follow-up of such studies is that, if the mass
of the central object M is determined using the NLR gas kinematics, the
presence of continuum  variability can, in principle, provide an
estimate for another key missing piece of information on AGN models,
the accretion rate \mdot. Using a very simple steady-state, irradiated
black-body approximation for the thermal structure of the accretion
disk, Peterson et al. \markcite{bpetal98} (1998) presented evidence of
a correlation between the product (M\,\mdot) and the time delay between
different continuum wavebands in NGC\,7469 (Wanders et al.
\markcite{iwatal97} 1997; Collier et al. \markcite{scetal98} 1998),
the only source where significant delays have been  detected. If
such correlation is found to hold for other nuclei, the two methods
above would provide totally independent estimates for each of the
quantities.  However, viscous dissipation time scales are too long to
reconcile with the observed continuum lags, which forces consideration
of X-ray irradiated or composite models (Sincell \& Krolik
\markcite{sk98} 1998; Collier et al.\markcite{scetal99} 1999).
Therefore, with an independent measure of the central object mass,
obtaining \mdot\ would be dependent on further development of the
accretion disk theory itself, and to obtain it using AGN continuum
variability would make it necessary to carry on  high sampling rate,
simultaneous multi-wavelength campaigns on other nearby active
galaxies.

\section{Summary} \label{sec_sumc}

We can summarize the main results of our study of both HST and
ground-based long-slit spectra of the inner and extended NLR of
\nfofo\ as follows:

\begin{enumerate}
\item By decomposing the \oiii\ line profile in multiple Gaussian
components we were able to trace the main kinematic component of the
ENLR across the nuclear region, connecting smoothly the emission gas
system with the large scale rotation defined by
\ion{H}{1}\ observations.

\item Individual clouds in the NLR (R $<$ 4\arcsec) are observed to be
kinematically disturbed by the interaction with the radio jet, but
underlying these perturbations the cloud system is moving in a pattern
compatible with disk rotation. High velocity components (up to
\mm\ 1000 \kms, relative to systemic) and broad (FWHM up to 1800 \kms) 
bases are detected in the \oiii\ profile of the brightest clouds. Such
regions are invariably at the edge of the radio knots, and this
association, together with the overall morphology of the velocity
field, lead us to propose that the main kinematic system in the inner
region of \nfofo\ is still rotation in the plane of the disk, {\it
disturbed but not defined} by the interaction with the radio jet and
the AGN emission.

\item Fitting a simple expression for planar rotation to the data, we
find that the ENLR gas (R $>$ 4\arcsec) presents a kinematic
behaviour consistent with and well represented by rotation in the
galactic disk, with characteristics similar to other normal spiral
systems. We obtain \pin\ = 21\arcdeg, and \ppo\ = 34\arcdeg\ --
43\arcdeg\ for the inclination to the line of sight and position angle
of the line of nodes of the disk, respectively.  The velocity field of
external knots at R \app\ 6\arcsec\ and 20\arcsec\ transverse to the
radial direction presents evidence of non-planar or non-circular
movements, probably associated with gas turbulence and streaming
motions along the bar.

\item Using the same projection angles as obtained for the ENLR, the
NLR emission component believed to represent the continuation of the
disk velocity field was also found to be consistent with  planar
rotation, although disturbed by the jet, as expected. However, while
the velocity field of the extended ENLR gas is dominated by the
potential of the galactic bulge, presenting a flat curve at large
distances, we find that the behaviour of the gas in the inner NLR is
best represented by a Keplerian-like potential, with the kinematics of
the gas up to 4\arcsec\ dominated by the \app\ \tento{9}\ \msol\ mass
concentration located within the turn-over radius of the rotation
curve, located at \app\ 0\arcsec.5. Our measurements inside the
turn-over radius imply that this is an extended distribution rather
than a point mass, and, if spherical symmetry is assumed, the innermost
observed velocity still not affected by the spatial PSF gives a 5
\vzs\ \tento{7} \msol\ mass concentrated within a radius of about 10
pc.

\end{enumerate}

\acknowledgments C.W.  to thanks the Space Telescope Science
Institute for the hospitality during the last two years, and
acknowledges the financial support from the Brazilian institution CNPq 
through a Post-doctoral fellowship, and from ESA. We also thank the 
referee, Dr. R.  Antonucci for his very thorough reading of this paper.

\newpage


\newpage



\figcaption[figure1.gif]{The slit positions of the ground-based
spectra of Table~\ref{tab_loggb} overlayed on the \oiii\ image of the
ENLR of \nfofo\ from P\'erez et al.\protect\markcite{epetal89} (1989).
\label{fig_slitsgb}}

\figcaption[figure2.gif]{[\ion{O}{3}] FOC f/96 image showing the
position of the spectrograph slit for the 1996 spectra. The actual
length of the slit is longer than represented. North is at the top and
East is to the left.  \label{fig_slits96}}

\figcaption[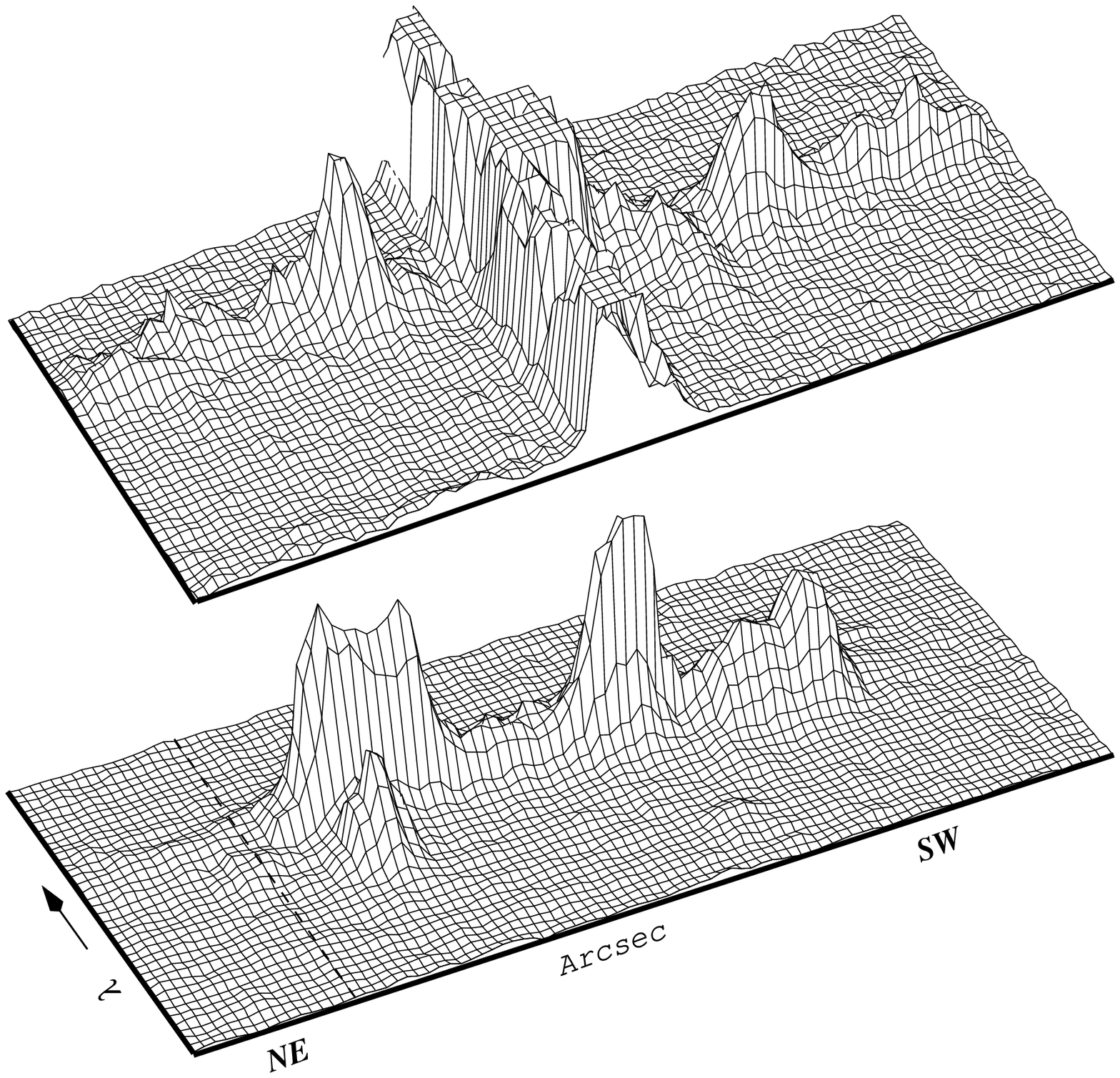]{Surface plots of a segment
of the nuclear (PA47\_1, top) and 0\arcsec.41 NW (PA47\_3, bottom) FOC
f/48 1996 spectra in the \oiii\ region.  Both regions are 2\arcsec.6 by
71 \AA\ (166 pc by 4250 \kms). \label{surface_fig}}

\figcaption[figure4.gif]{[\ion{O}{3}] FOC f/96 image showing the
position of the spectrograph slit for the 1995 spectra. Orientation as
in Figure~\ref{fig_slits96}.  \label{fig_slits95}}

\figcaption[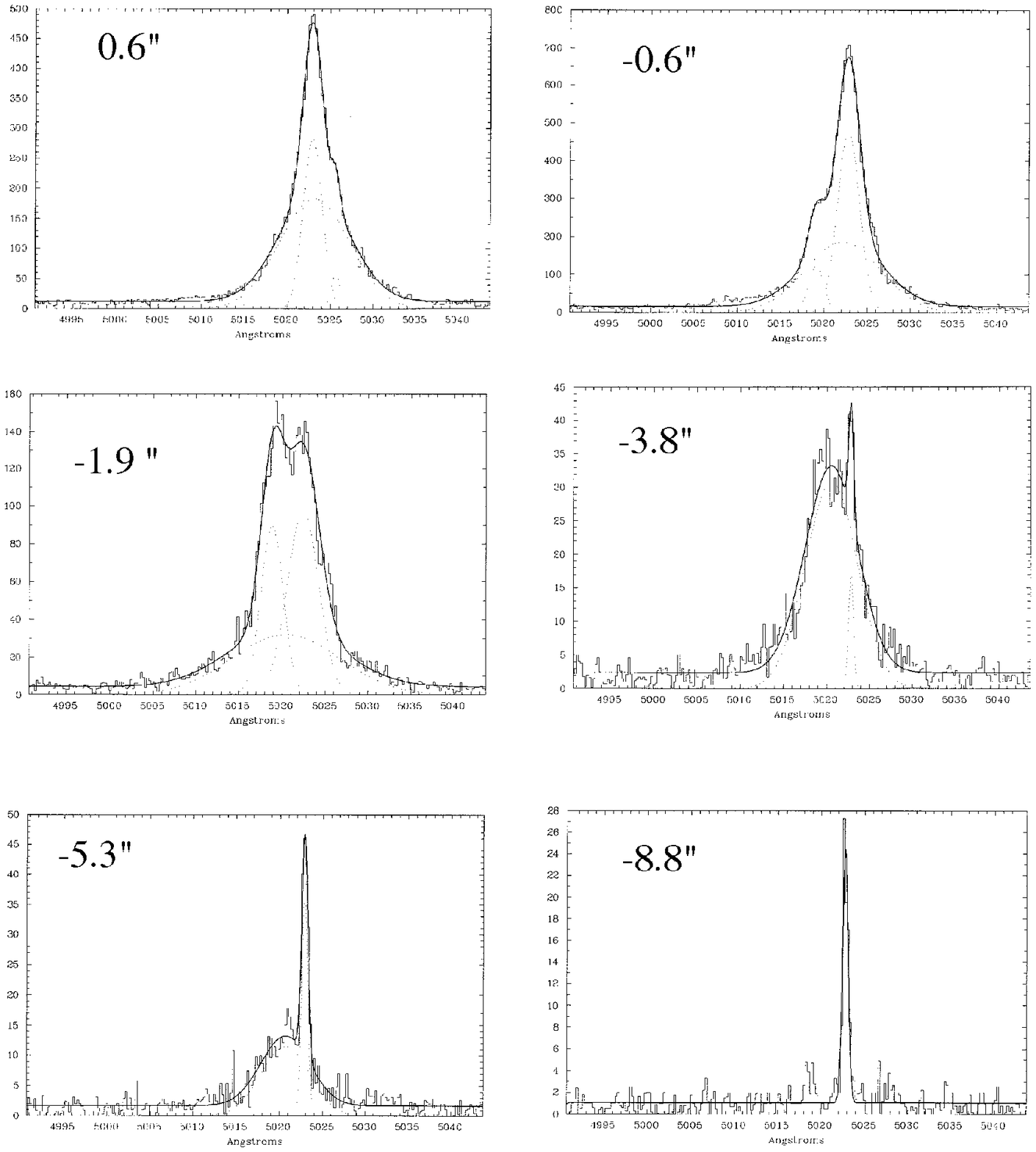]{Sample of INT \oiii\ line profiles and
Gaussian fitting. From top left to bottom right: (a) 0\arcsec.6 NW of
the nucleus; (b) 0\arcsec.6 SW; (c) 1\arcsec.9 SW; (d) 3\arcsec.8 SW;
(e) 5\arcsec.3 SW; and (f) 8\arcsec.8 SW. Notice how the narrow
extended component can be traced into the inner 5\arcsec\ of the
NLR.\label{fig_gaussgb} }

\figcaption[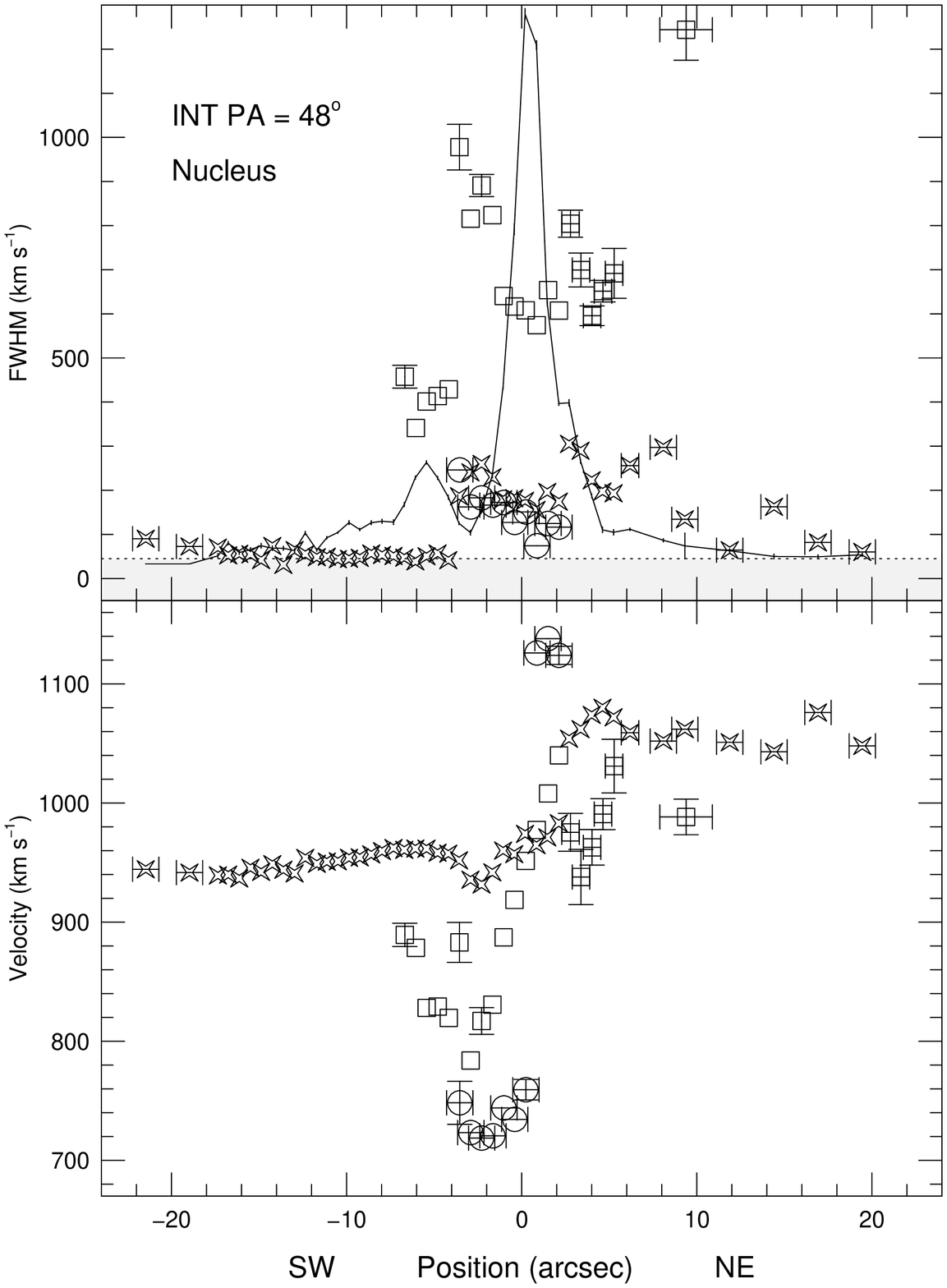]{Result of the Gaussian decomposition
procedure for the \oiii\ line of the INT PA48 spectrum. The velocities
derived from the central wavelength of the individual components are
shown in the bottom panel. The top panel presents the corresponding
FWHM, and the solid line is the brightness profile of the narrow
component. The data are plotted as a function of the distance to the
center of the slit. For this and the other similar figures in the
paper, error bars smaller than the point size are not plotted. (Stars)
main narrow component; (square) broad component; (open circle)
high-velocity narrow component. \label{fig_vobsgb} }

\figcaption[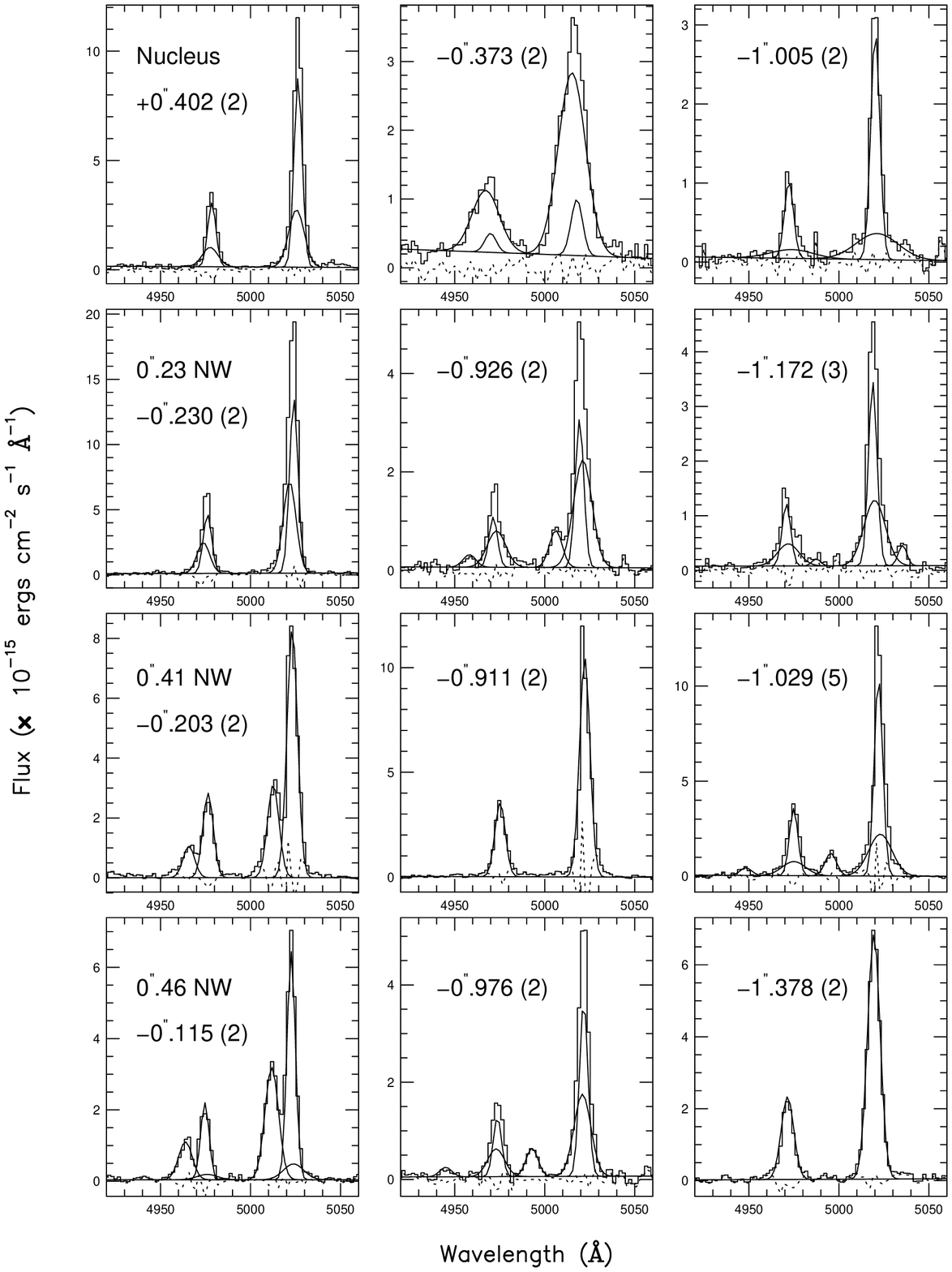]{ Gaussian fitting to a sample of \oiiit
4959,5007 line profiles from the FOC f/48 1996 spectra.  The spectra
are identified by their distance in arc seconds to the center of the
corresponding slit, negative values to the SW. The number after the
distance corresponds to how many columns in the image were coadded.
The histogram is the data, the thin lines the individual components and
underlying continuum, and the dotted line represents the residuals of
the fit.  \label{fig_gaussfit}}

\figcaption[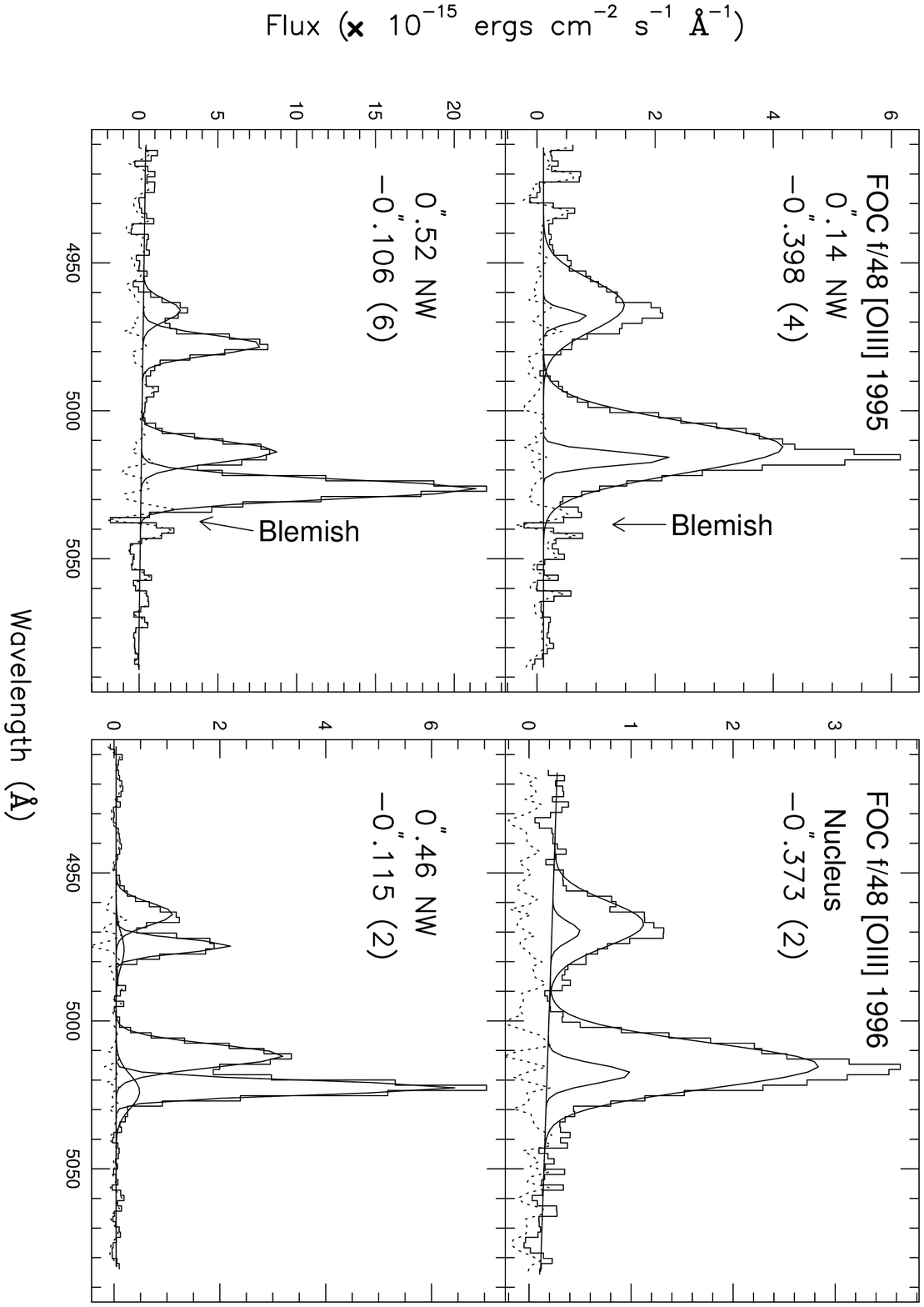]{Same as Figure~\ref{fig_gaussfit} for two
positions on the 1995 FOC f/48 spectra (left), and the regions closest
to them in the 1996 data (right)  \label{fig_gauss95} }

\figcaption[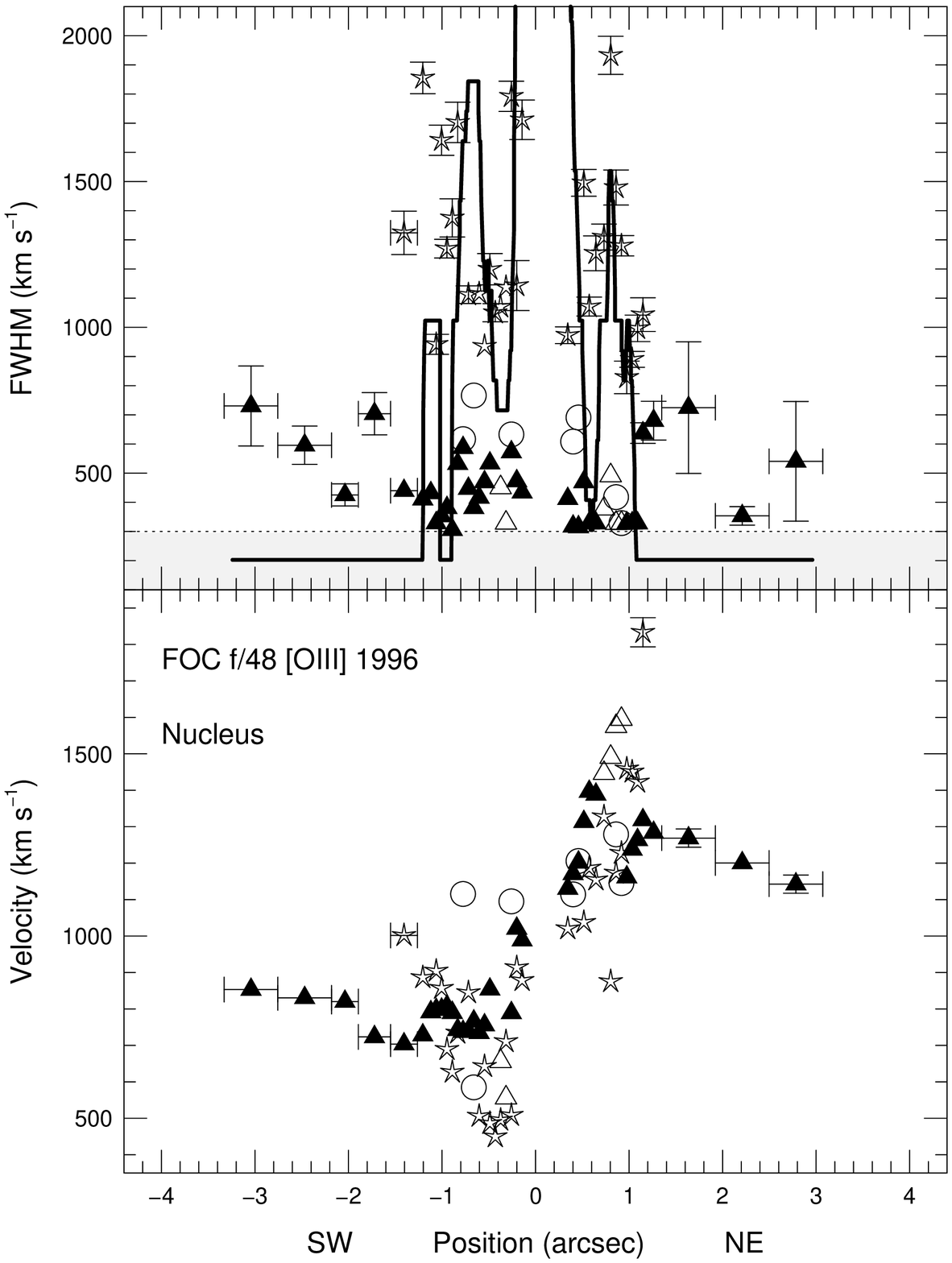,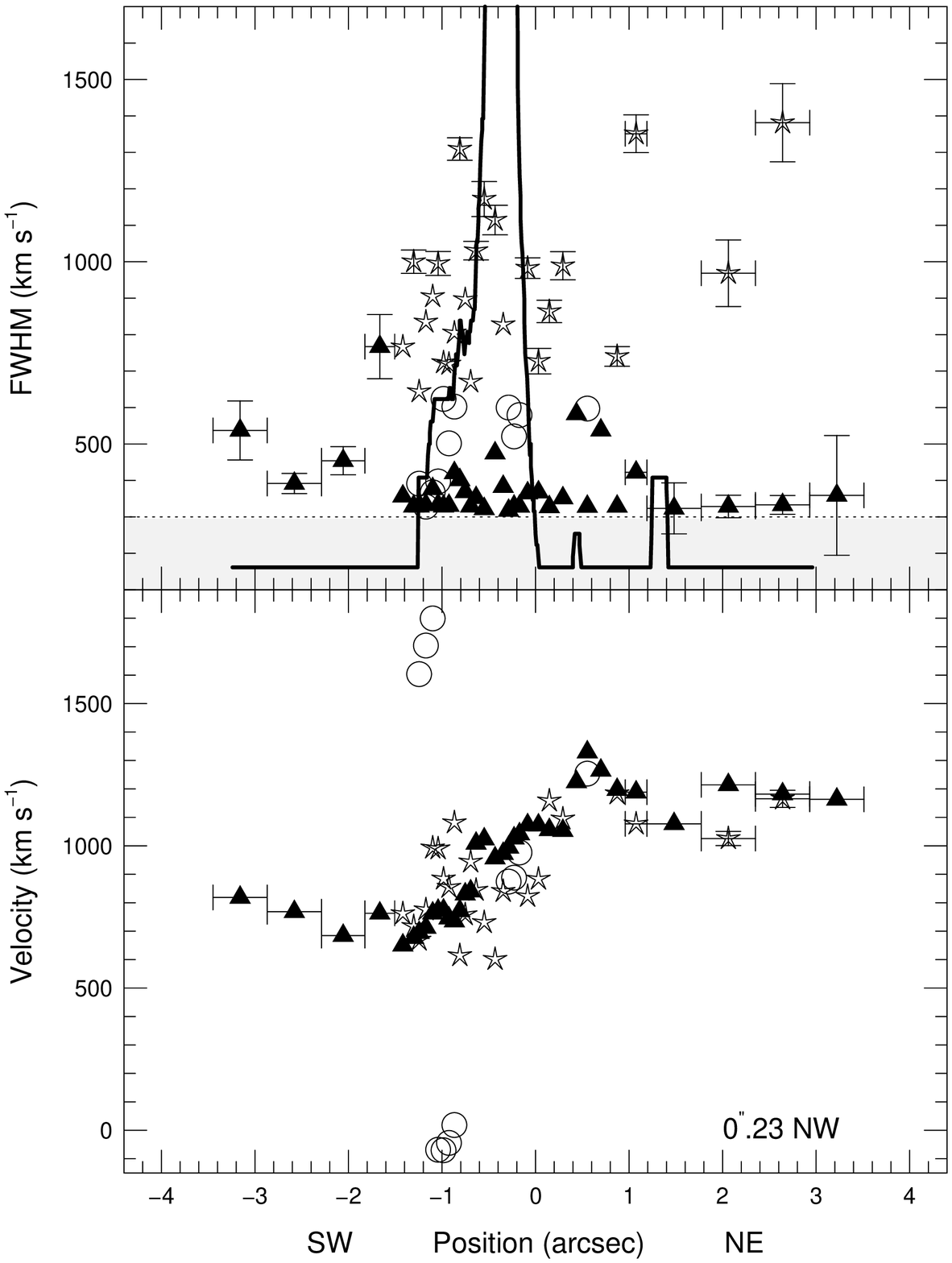,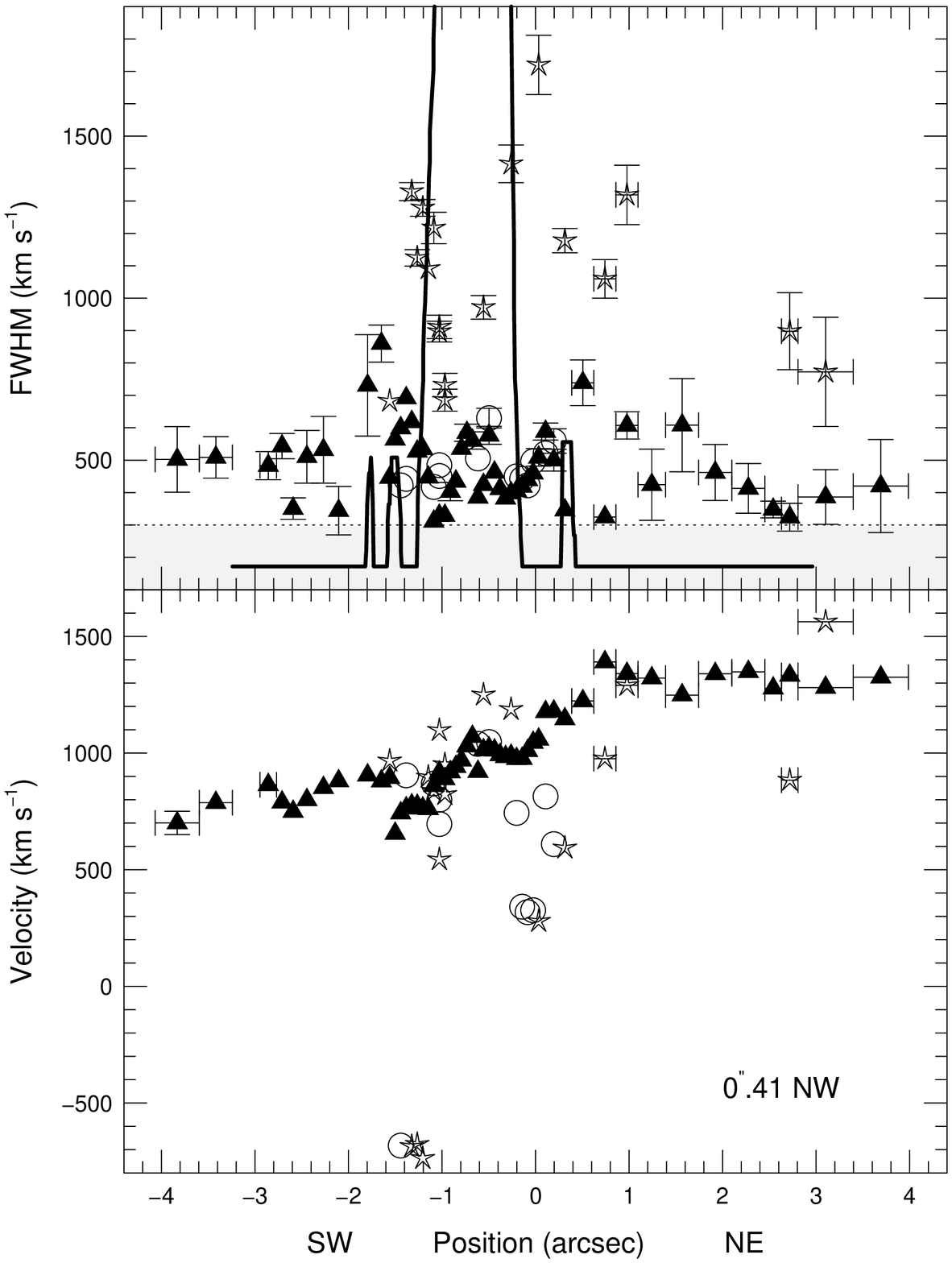,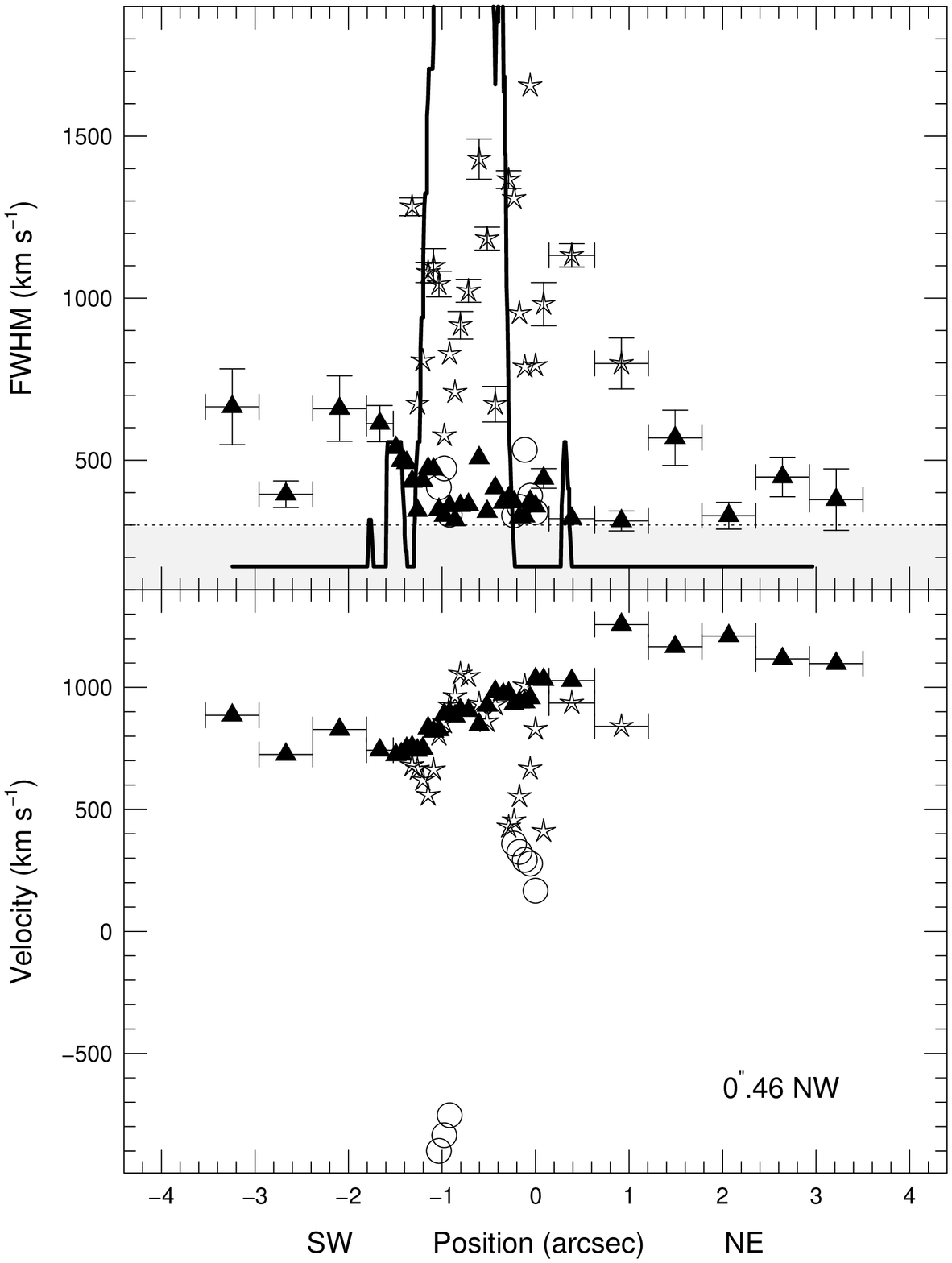]{Result
of the Gaussian decomposition procedure for the \oiii\ line of the 1996
FOC f/48 spectra: (a) Nucleus, (b) 0\arcsec.23 NW, (c) 0\arcsec.41 NW,
and (d) 0\arcsec.46 NW. The velocities derived from the central
wavelength of the individual components are shown in the bottom panels.
The top panels present the corresponding FWHM superposed to the
VLA$+$Merlin 5 GHz brightness profile at each slit position. Both are
plotted as a function of the distance to the center of the slit.
(Filled triangle) main narrow component; (open circle) narrow secondary
component; (stars) broad component. \label{fig_vobso3} }

\figcaption[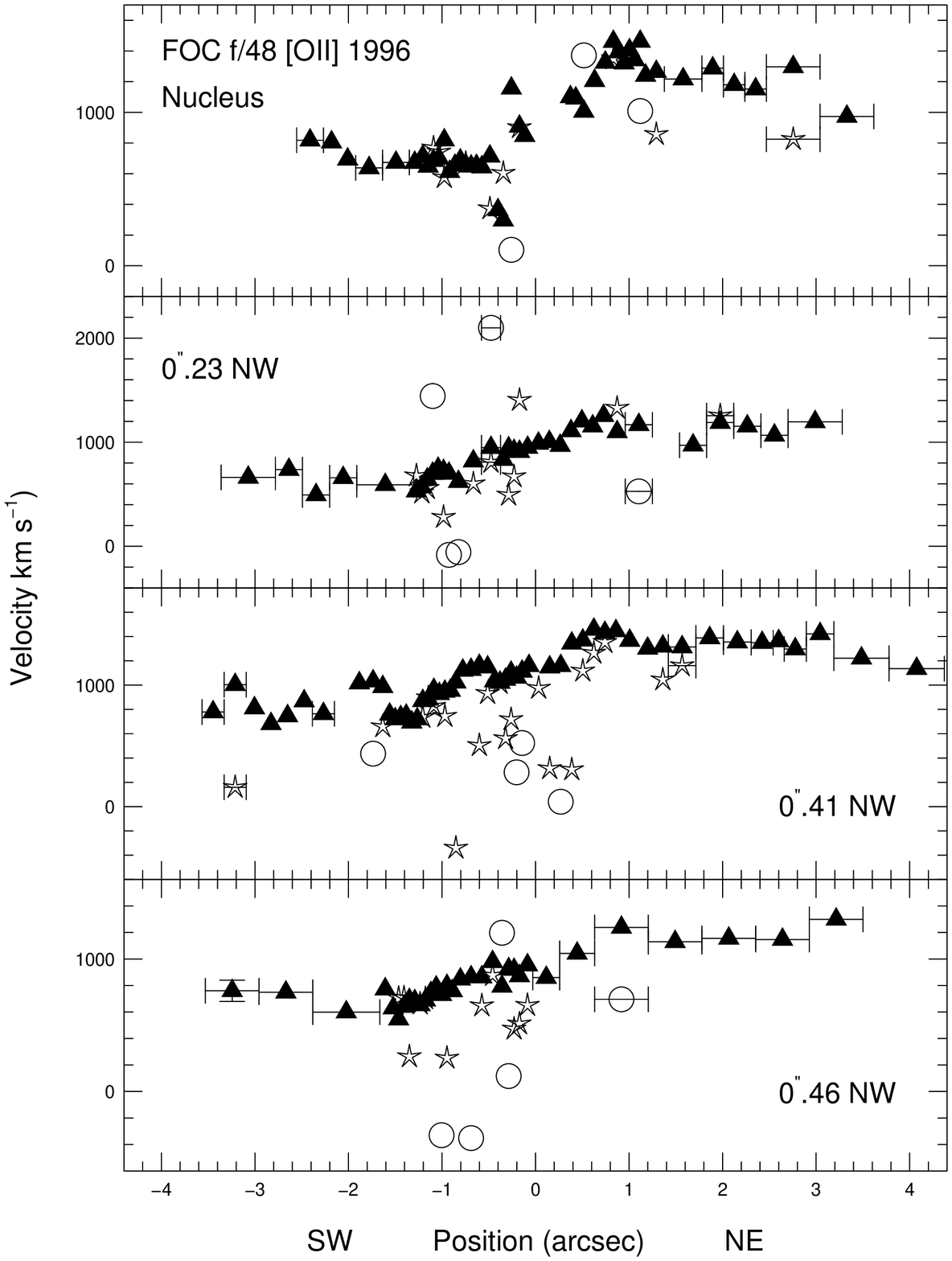]{Rotation curves derived from the Gaussian
decomposition of the 1996 FOC f/48 \oii\ line. From top to bottom: (a)
Nucleus, (b) 0\arcsec.23 NW, (c) 0\arcsec.41 NW, and (d) 0\arcsec.46
NW. Symbols as in Figure~\ref{fig_vobso3}. \label{fig_vobso2} }

\figcaption[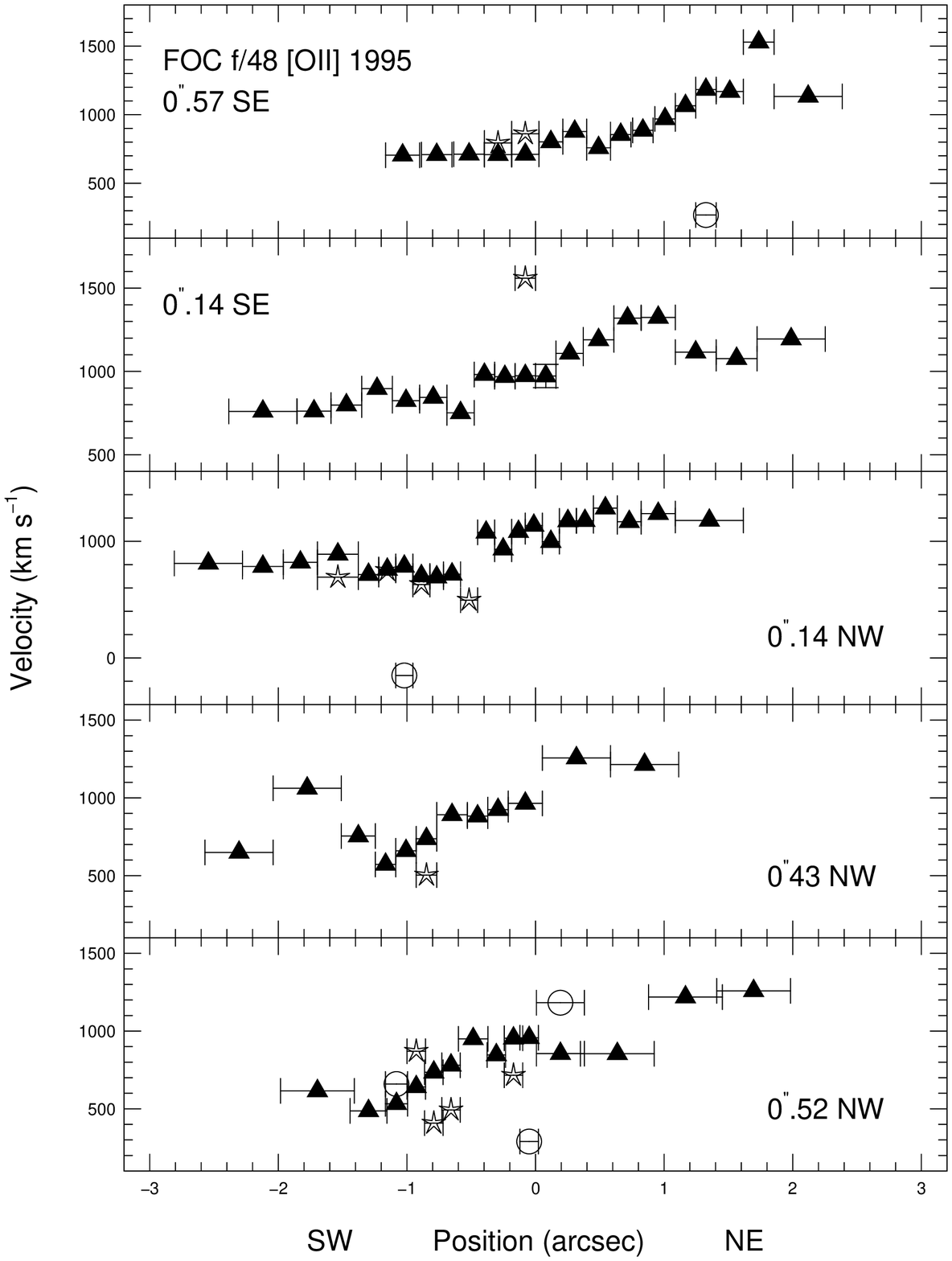]{Same as in Figure~\ref{fig_vobso2} for the
1995 FOC f/48 \oii\ line. From top to bottom: (a) 0\arcsec.57 SE; (b)
0\arcsec.14 SE; (c) 0\arcsec.14 NW;, (d) 0\arcsec.43 NW; (e)
0\arcsec.52 NW. Symbols as in Figure~\ref{fig_vobso3}.
\label{fig_vobso2a} }

\figcaption[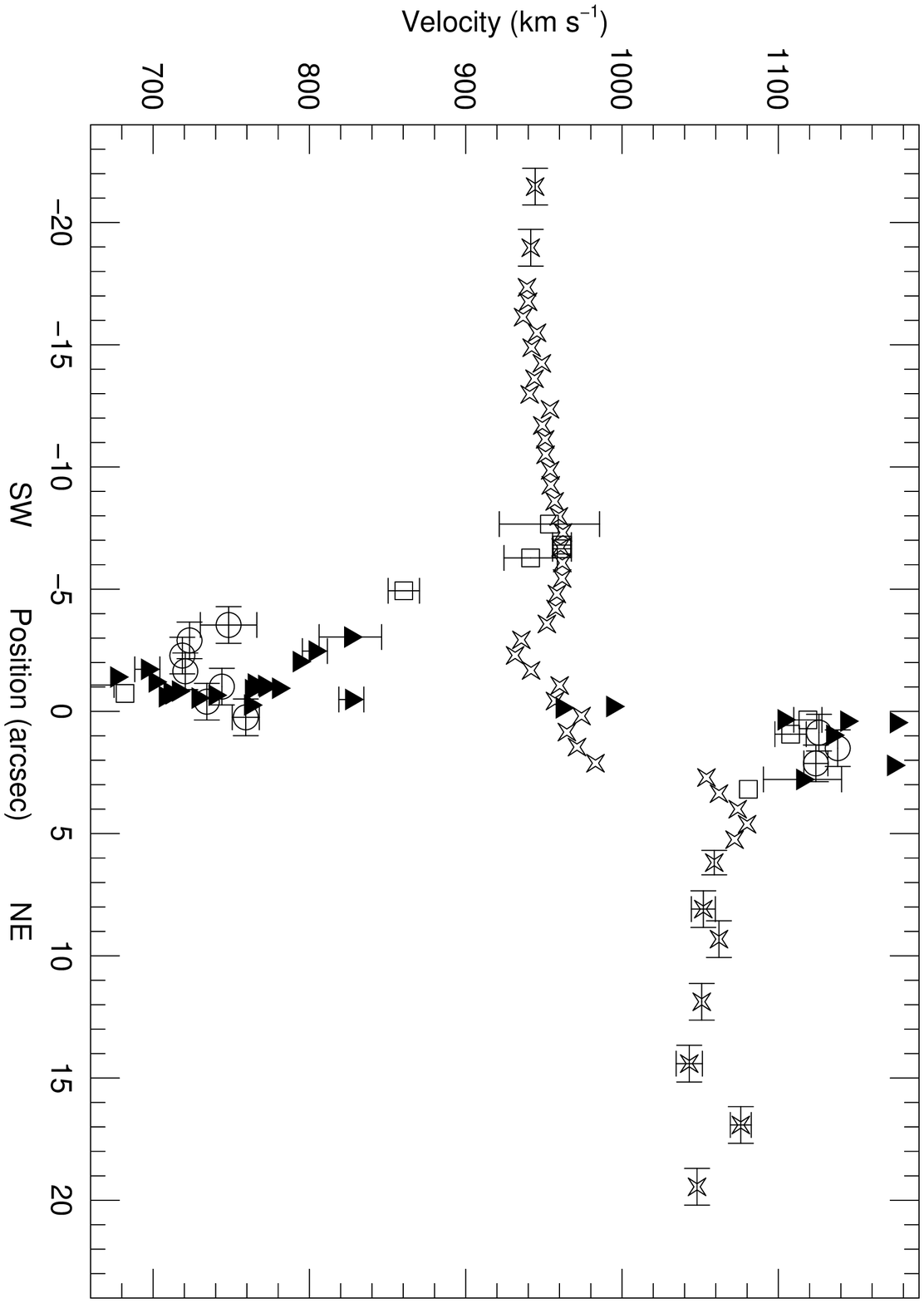]{Comparison between the ground-based ENLR
rotation curve (PA48 position; 1st component, stars; 3rd component,
open circles), with the FOC f/48 data (PA47\_1; filled triangles) and
the STIS slitless data from Hutchings et al. (1998) (open squares) corresponding to clouds closer to PA=48\arcdeg. \label{fig_pa48all_fit08} }

\figcaption[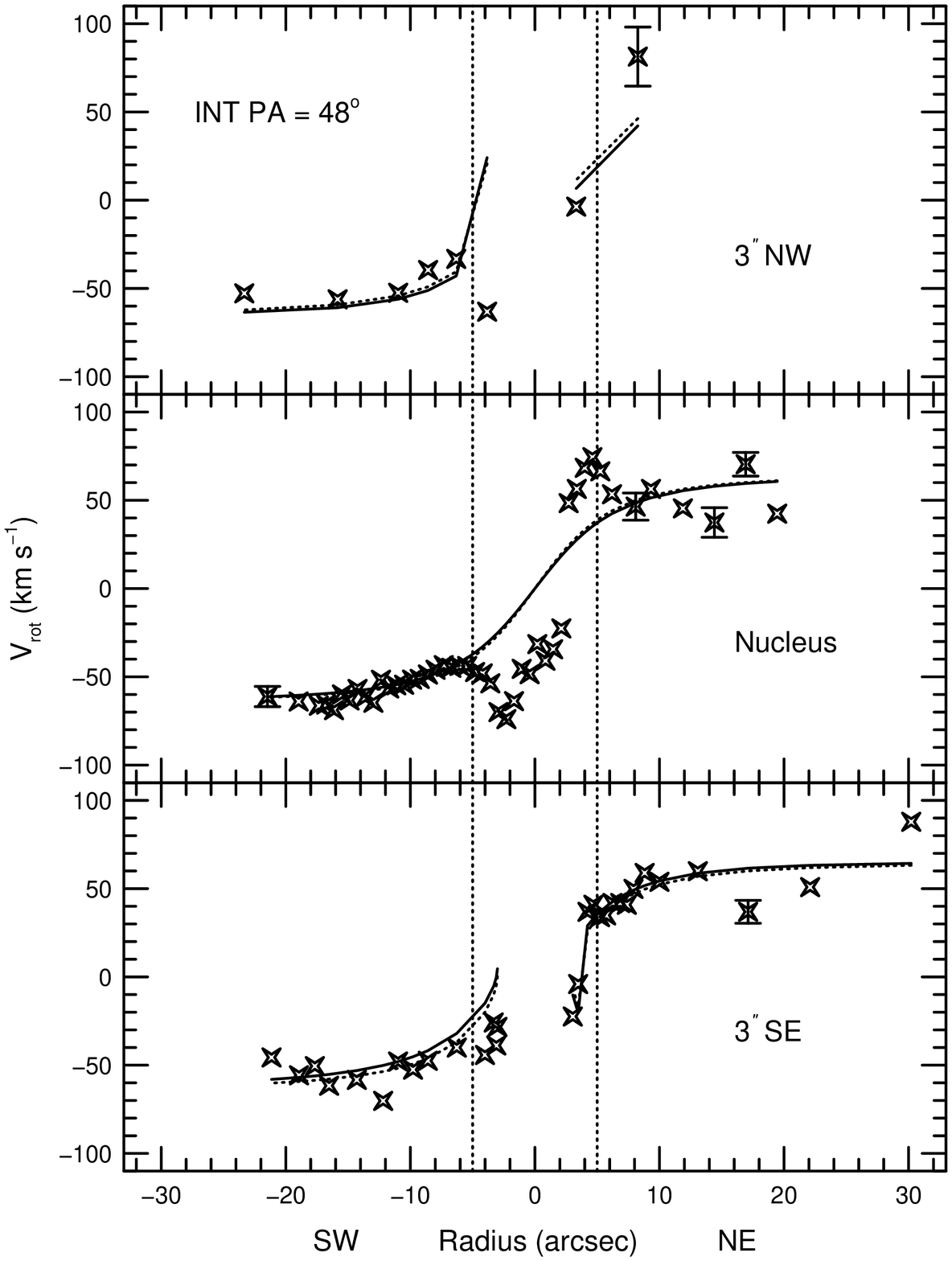]{Comparison of the observed radial
velocities from the INT (PA$=$48\arcdeg) data with the best fitting
circular rotation curve from Eq.~\ref{eq:rot}, corresponding to Models
A (solid line) and B (dotted line) on Table~\ref{tab_modgb}. The
vertical lines mark the region excluded from the fit due to the
interaction of the emission gas with the radio jet. \label{fig_fitgb} }

\figcaption[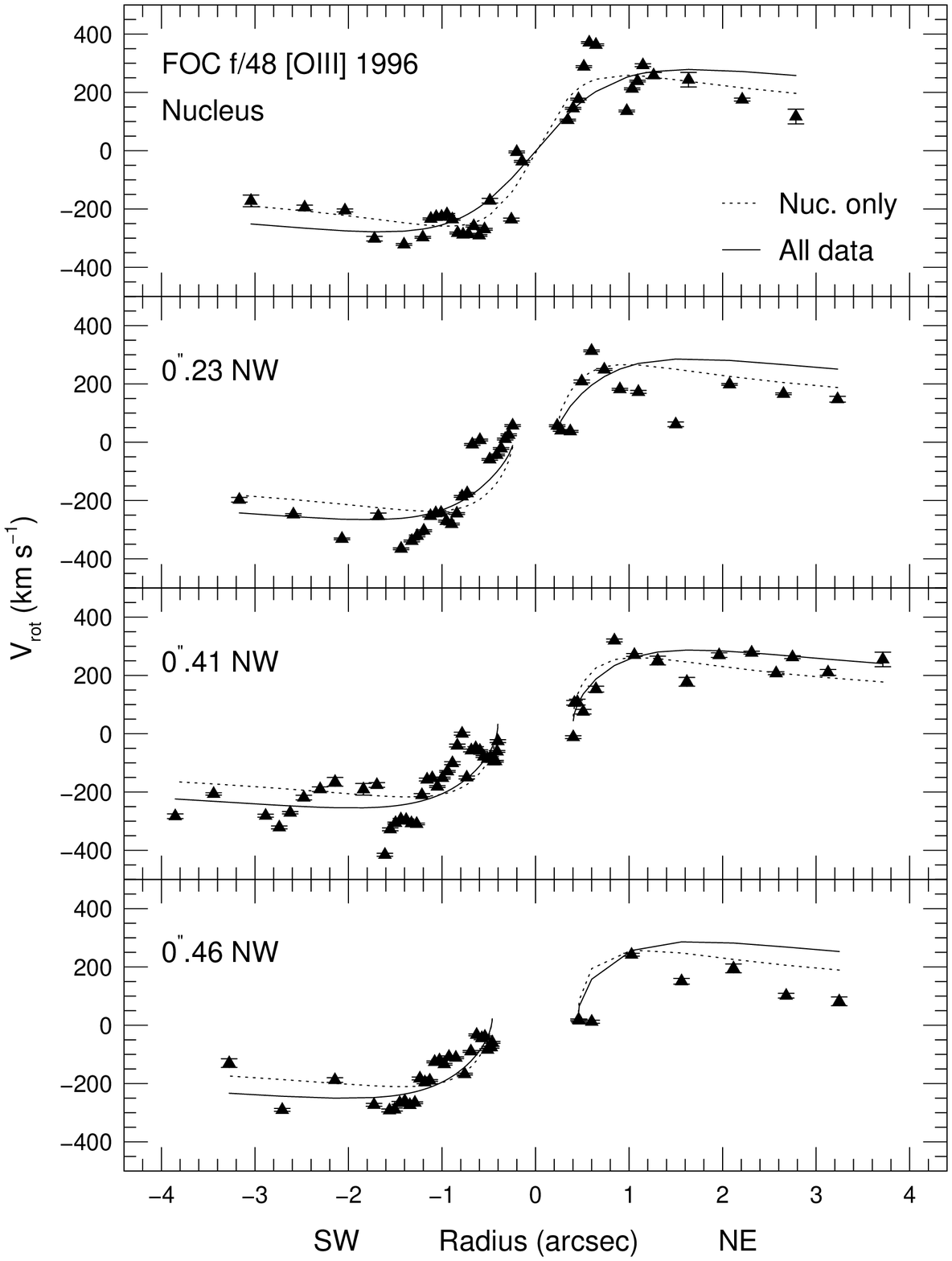]{Comparison between the 1996 FOC f/48
\oiii\ data set and the best fitting circular rotation curve. The
dotted line corresponds to the fit to the nuclear slit position only,
and the solid line to the fit obtained using all the four slit
positions simultaneously. Notice that the distance scale is now
relative to the nucleus.  \label{fig_fitst1} }

\figcaption[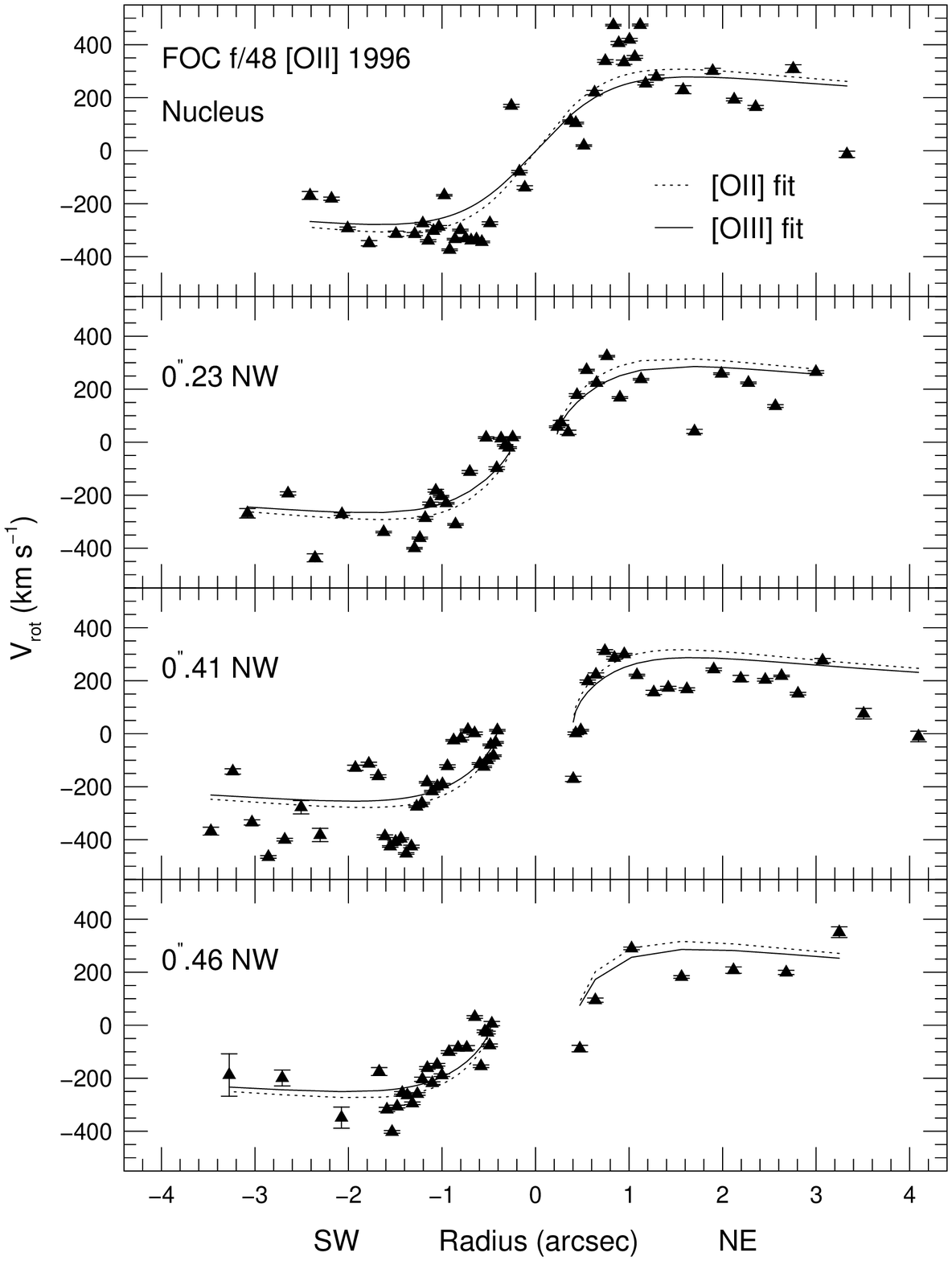]{Same as Figure~\ref{fig_fitst1} for the
1996 FOC f/48 \oii\ data set. The solid line corresponds to the fit
obtained with the \oiii\ data set (all positions), and the dotted line
to the fit obtained using the \oii\ emission from the four slit
positions simultaneously. \label{fig_fitst2a}  }

\figcaption[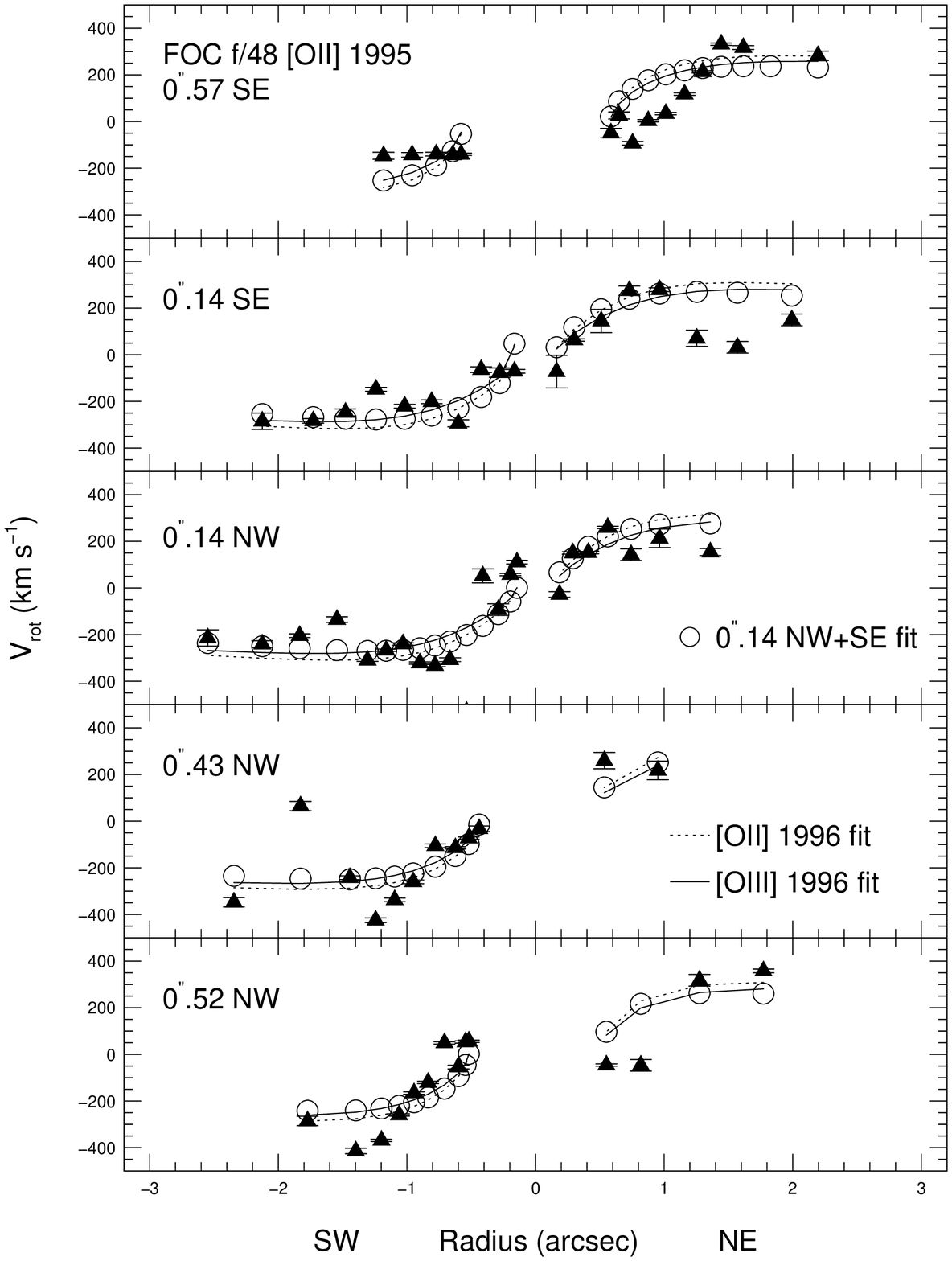]{Same as Figure~\ref{fig_fitst1} for
the 1995 FOC f/48 \oii\ data set. Here the solid line and dotted lines
correspond to the fits obtained with the 1996 \oiii\  and \oii\ data
sets (all positions), respectively,  and the open circles represent the
fit obtained using the 1995 \oii\ emission from the two innermost
(0\arcsec.14 NW and 0\arcsec.14 SE) slit positions. \label{fig_fitst2b}
}

\figcaption[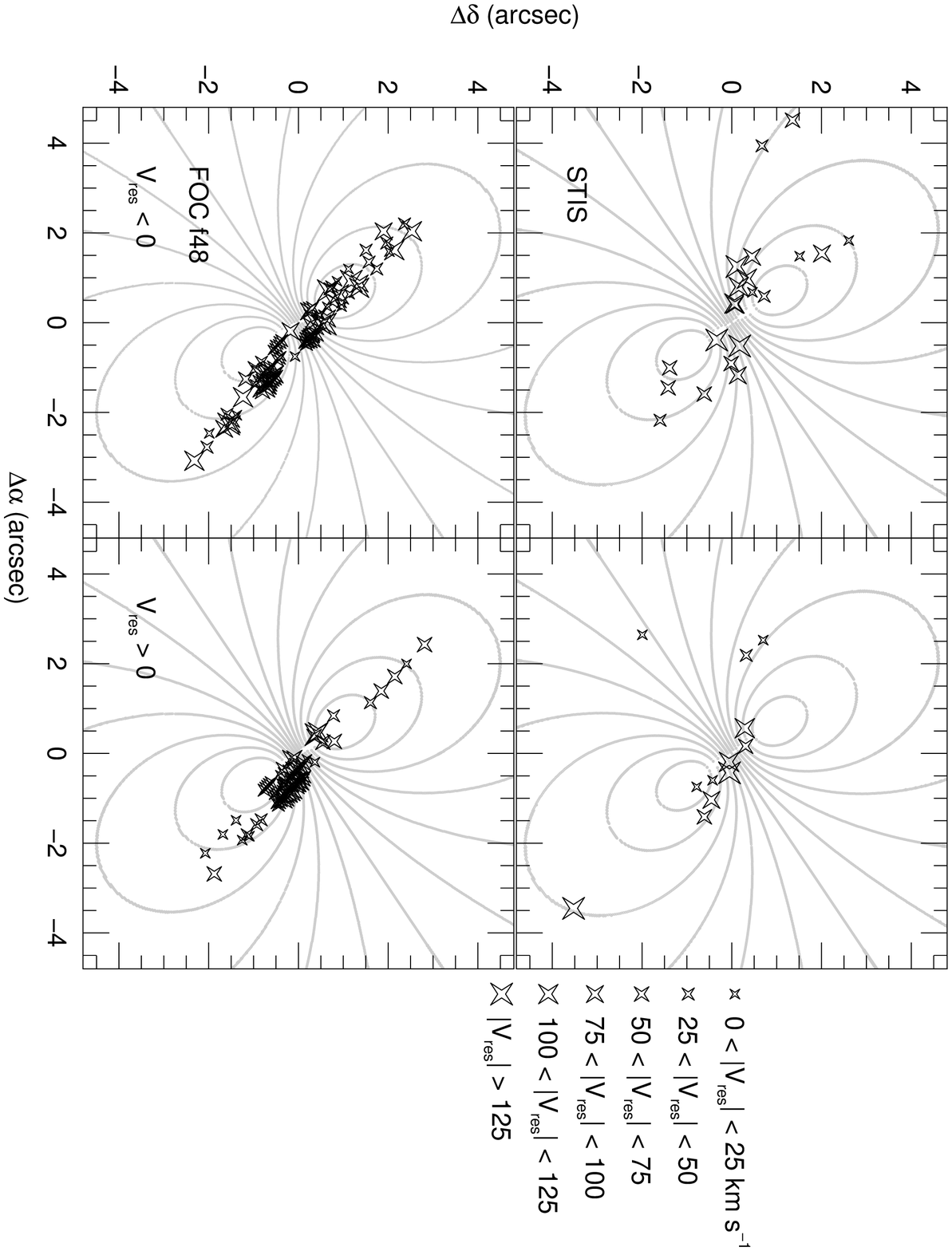]{Bidimensional representations of the
rotation model for the STIS data  (upper panels) and the \oiii\ Nuc.
model of Table~\ref{tab_modst} for the FOC f/48 data (bottom). North is
up; East to the left. The velocity contours are 0, 40, 80, 120, 160,
200, 240 \kms\ to the NE and the negative equivalents to the SW. The
points are the residuals, defined as $V_{res} = V_{rot} - V_{model}$,
negative  on the left, and positive on the right panels. \label{fig_spiderd} }

\newpage


\begin{deluxetable}{lccccc}
\tablecolumns{6}
\tablewidth{0 pt}
\tablecaption{Journal of Observations -- Ground Based Data \label{tab_loggb}}
\tablehead{\colhead{} & \colhead{PA} & \colhead{Offset} & \colhead{Exp. time} & 
\colhead{Slit} 
& \colhead{Seeing} \\    
\colhead{Name} & \colhead{(Degrees)} & \colhead{from Nucleus} & \colhead{(s)} & 
\colhead{Width} & \colhead{(arcsec)} }
\startdata
5\,NE                    & 138 & 5\arcsec NE   & 1000    & 0\arcsec.6\phn & 2.7 \nl
2.5\,NE                  &     & 2\arcsec.5 NE & \phn510 & 0\arcsec.65    & 1.7 \nl
PA138                    &     & Nucleus       & 1050    & 0\arcsec.62    & 2.2 \nl
2.5\,SW\tablenotemark{a} &     & 2\arcsec.5 SW & \phn500 & 0\arcsec.65    & 1.7 \nl
5\,SW                    &     & 5\arcsec SW   & 1000    & 0\arcsec.6\phn & 2.7  \nl
10\,SW                   &     & 10\arcsec SW  & 1000    &                &      \nl
20\,SW                   &     & 20\arcsec SW  & 1000    &                &      \nl
3\,NW                    & 48  & 3\arcsec NW   & 1000    &                &      \nl
PA48                     &     & Nucleus       & 1000    & 0\arcsec.22    & 2.6  \nl
3\,SE                    &     & 3\arcsec SE   & 1000    & 0\arcsec.6\phn & 2.7  \nl
\enddata
\tablenotetext{a}{Spectrum obtained with a 0.6 neutral density filter.}
\end{deluxetable}

\newpage

\begin{deluxetable}{lccclc}
\tablecolumns{6}
\tablewidth{466 pt}
\tablecaption{Journal of Observations -- HST Data \label{tab_logst}}
\tablehead{\colhead{Name} & \colhead{Dataset}& \colhead{Date} & 
           \colhead{Exp.\,time(s)} & \colhead{Format} & \colhead{Comments}} 
\startdata
\cutinhead{Science Data}
NGC\,4151 & X2PJ010CP & 1995 Jul 11 &\phn934  & 1024\vzs 256 & PA40\_1 
0\arcsec.57 SE \nl
          & X2PJ010DT &             &\phn934  & 1024\vzs 256 & PA40\_2 0\arcsec.14 SE 
\nl
          & X2PJ010IT &             &\phn598  & 1024\vzs 256 & PA40\_3 0\arcsec.14 NW 
\nl
          & X2PJ010JT &             &\phn598  & 1024\vzs 256 & PA40\_4 0\arcsec.43 NW 
\nl
          & X2PJ010KT &             &\phn598  & 1024\vzs 256 & PA40\_5 0\arcsec.52 NW 
\nl
NGC\,4151 & X38I0108T & 1996 Jul 03 &\phn697  & 1024\vzs 512 & PA47\_1 
Nucleus \nl
          & X38I0109T &             &\phn697  & 1024\vzs 512 & PA47\_2 0\arcsec.23 NW 
\nl
          & X38I0102T &             & 1247    & 1024\vzs 512 & PA47\_3 0\arcsec.41 NW \nl
          & X38I010AT &             &\phn697  & 1024\vzs 512 & PA47\_4 0\arcsec.46 NW 
\nl
\cutinhead{Calibration Frames}
NGC\,4151 & X2PJ010GT & 1995 Jul 11 &\phn600  & 1024\vzs 512z & internal flat 
\nl
          & X38I0106T & 1996 Jul 03 &\phn600  & 1024\vzs 512  & internal flat \nl
NGC\,6543 & X3BD0102T & 1996 Sep 10 &\phn682  & 1024\vzs 512  & Planetary 
Neb. \nl
          & X3BD0103T &             &\phn717  & 1024\vzs 256  & Planetary Neb. \nl
          & X3BD0104T &             &\phn600  & 1024\vzs 512z & internal flat \nl 
          & X3BD0105T &             &\phn500  & 1024\vzs 512  & internal flat \nl
47 Tuc.   & X34I0108T & 1996 Apr 04 &\phn477  & 1024\vzs 256z & Globular Cl. \nl
          & X34I0109T &             &\phn600  & 1024\vzs 256z & internal flat \nl
LDS\,749b & X3L8020BT & 1996 Dec 27 &\phn497  & 1024\vzs 512  & Flux Standard 
\nl
          & X3L8020AT &             &\phn800  & 1024\vzs 512  & internal flat \nl
\enddata
\end{deluxetable}

\newpage

\begin{deluxetable}{lccc}
\tablecolumns{4}
\tablewidth{0 pt}
\tablecaption{Model Fitting - Ground Based \tablenotemark{a} \label{tab_modgb}}
\tablehead{\colhead{Parameter} & \colhead{Range} & \colhead{Model A} & 
\colhead{Model B}}    
\startdata
\pa\ (\kms)    & 170 -- 186  & \phd181          & \phd182     \nl
\pp            & 1.0         & \phn\phn1.0  & \phn\phn1.0 \nl
\pco\ (\arcsec)& 6.4 --7.4   & \phn\phn7.1  & \phn\phn6.6   \nl
\ppo\ (\arcdeg)& 31 -- 48    & \phn33.9     & \phn43.7     \nl
\pin\ (\arcdeg)& 18 -- 23    & \phn21.8     & \phn20.9  \nl
\enddata
\tablenotetext{a}{PA=48\arcdeg data}
\end{deluxetable}

\newpage

\begin{deluxetable}{lccccccc}
\tablecolumns{8}
\tablewidth{0 pt}
\tablecaption{Model Fitting - HST FOC f/48 \label{tab_modst}\tablenotemark{a}}
\tablehead{\colhead{} & \multicolumn{3}{c}{1996 [\ion{O}{3}]} & 
\multicolumn{2}{c}{1996 [\ion{O}{2}]} &  \multicolumn{2}{c}{1995 [\ion{O}{2}]}\\
\colhead{Parameter} & \colhead{Range} & \colhead{fit} & \colhead{Nuc.} & 
\colhead{Range} & \colhead{fit} & \colhead{Range} & \colhead{fit} }   
\startdata
\pa\ (\kms)    & 1382 -- 1412 & 1399\phd  &  986\phn\phd  & 1434 -- 1508 & 1478\phd 
& 1133 -- 1175 & 1157\phd \nl
\pp           & 1.499        & 1.499 & 1.499 & 1.499 & 1.499 & 1.498 - 1.50 & 1.498  \nl
\pco\ (\arcsec)& 1.16 -- 1.19 & 1.18\phn & 0.68\phn & 1.04 -- 1.10  & 1.07\phn &  
0.84 -- 0.88 & 0.86\phn \nl
\enddata
\tablenotetext{a}{Using \ppo = 33\arcdeg.9 and \pin = 21\arcdeg.}
\end{deluxetable}

\clearpage


\end{document}